\documentclass[sigconf]{acmart}
\renewcommand\footnotetextcopyrightpermission[1]{}

\usepackage{amssymb}
\usepackage{amsfonts}
\usepackage{algorithmic}
\usepackage{graphicx}
\usepackage{xcolor}
\usepackage{makecell}
\usepackage{booktabs} 
\usepackage{amsmath}
\usepackage[normalem]{ulem}
\usepackage{enumitem}
\usepackage{multirow}
\usepackage[nomargin,inline,marginclue,draft]{fixme}
\usepackage{balance}
\usepackage{changepage}
\usepackage{bm}
\usepackage{hyperref}
\usepackage{setspace}
\usepackage{mathrsfs}
\usepackage{ulem}
\usepackage{verbatim}
\usepackage{diagbox}
\usepackage{pdftexcmds}
\usepackage{catchfile}
\usepackage{ifluatex}
\usepackage{ifplatform}
\usepackage{graphicx}
\usepackage{subfigure}

\newlength\savedwidth

\author{Chang Liu, Chen Gao, Depeng Jin, Yong Li}
\affiliation{
\institution{Department of Electronic Engineering, Tsinghua University}
	\country{}
}
\email{liyong07@tsinghua.edu.cn}
		
\begin{document}
	\title{Improving Location Recommendation\\ with Urban Knowledge Graph}
	\begin{abstract}
	
Location recommendation is defined as to recommend locations (POIs) to users in location-based services. The existing data-driving approaches of location recommendation suffer from the limitation of the implicit modeling of the geographical factor, which may lead to sub-optimal recommendation results. In this work, we address this problem by introducing knowledge-driven solutions. Specifically, we first construct the Urban Knowledge Graph (UrbanKG) with geographical information and functional information of POIs. On the other side, there exist a fact that the geographical factor not only characterizes POIs but also affects user-POI interactions. To address it, we propose a novel method named UKGC. We first conduct information propagation on two sub-graphs to learn the representations of POIs and users. We then fuse two parts of representations by counterfactual learning for the final prediction. Extensive experiments on two real-world datasets verify that our method can outperform the state-of-the-art methods.
	
    \end{abstract}
	
	\maketitle
	
	\section{Introduction}\label{sec::intro}

Location-based Social Networks (LBSNs) enable users to share their locations, check-ins, photos and comments via the Internet, which produces a large amount of data and facilitates the development of the Point-of-Interest recommendation. 
Besides the collaborative filtering methods~\cite{he2020lightGCN,Wang2019ngcf} that learn users' preferences solely on the user-POI interaction data, 
various works leverage different kinds of side information to enhance the POI recommendation, such as geographical factors~\cite{wang2018geoPOI}, social networks~\cite{zhang2019kean}, categories~\cite{lin2018cate}, etc.
Nevertheless, due to the lack of structural data, these works are still limited by the black-box properly of machine learning models, which leads to sub-optimal results.

Overall speaking, the interactions between users and POIs are not entirely determined by the interests of users and the functional attributes of POIs.
More precisely, a user's visit to a specific POI does not only mean that the POI's functional attributes are entirely matched to the requirements of users, but may indicate that the geographical distance between the user and the POI is relatively close.
It can be regarded as a kind of \textit{geographical bias} in the observed user-POI interaction data.
Therefore, from the causal inference perspective~\cite{pearl2018book}, geographical factor not only characterizes POIs (the \textit{cause}) but also affects user-POI interactions (the \textit{effect}), and thus serves as a \textit{confounder}.
This will not only make it difficult to accurately match users' interests but drastically damage the performance of recommender systems. Geographical factors are important attributes of POIs intrinsically.
Therefore, for a better learning of user interests, the recommendation model should ensure the geographical factor only affects the attributes of POIs (indirect effect) rather than directly affects the interactions.

To address the limitation of existing works and weaken the influence of the bias caused by the geographical confounder, in this paper, we first build large-scale
Urban Knowledge Graphs (UrbanKG) in two big cities, which provide rich knowledge on both geographical and functional aspects. After that, 
we divide the UrbanKG into two subgraphs, called \textit{geographical graph} and \textit{functional graph} according to the attributes of relations.
We then design a graph neural network-based method that both learns from two subgraphs derived from UrbanKG and the user-POI interaction graph. Disentangled embeddings of POIs and users are learned for capturing their geographical and functional representations.
Furthermore, we propose the causal graph based on which we further propose a counterfactual learning method for the final recommendation.

The main contributions of this paper can be summarized as follows:
\begin{itemize}[leftmargin=*,partopsep=0pt,topsep=0pt]
\setlength{\itemsep}{0pt}
\setlength{\parsep}{0pt}
\setlength{\parskip}{0pt}
    \item To our best knowledge, we are the first to build large-scale UrbanKGs for POI recommendation. We also collect user-POI interactions that involve POIs overlapped with the UrbanKG. We hope that these datasets can benefit the community.
    \item We propose the UKGC model that can learn from both the interaction graph and knowledge graphs well with disentangled embeddings. We also propose the counterfactual learning for recommendation, which can eliminate the geographical bias and model users' interests accurately.
    \item We conduct extensive experiments to evaluate the performance and results show superior performances of our method. The results also demonstrate the effectiveness and utility of our UrbanKG. Further experiments also validate the rationality of each component of our UKGC model.
\end{itemize}
\textbf{Notations and Problem Formulation.}
Let $\mathcal{U}$ be the set of all users and $\mathcal{P}$ be the set of all POIs. Define the set of user-POI interactions as $\mathcal{O^+}=\{(u,p)|u\in \mathcal{U}, p\in \mathcal{P}\}$, in which $(u,p)$ means that the user $u$ has visited the POI $p$ in history.
We formulate the problem to be addressed in this work as follows,
\begin{itemize}[leftmargin=*,partopsep=0pt,topsep=0pt]
    \setlength{\itemsep}{0pt}
    \setlength{\parsep}{0pt}
    \setlength{\parskip}{0pt}
    \item \textbf{Input:} the user-POI interaction dataset $\mathcal{O}^+$ and the urban knowledge graph $\mathcal{G}$, which we will describe later.
    \item \textbf{Output:} a function that can predict the probability of interaction between the user $u$ and the POI $p$, denoted as $\hat{y}_{u,p}$.
\end{itemize}
	
\section{Urban Knowledge Graph Overview}\label{sec::UKG}
Earlier works about POI recommendations only used the historical interaction information to recommend similar POIs to users with similar profiles based on their preference, i.e., user-based collaborative filtering or recommend POIs to users based on the similarity between POIs, i.e., content-based collaborative filtering. However, due to the limited interaction data between users and POIs and unbalanced proportion of positive and negative samples, it will be difficult to learn the representations of users and POIs well. In recent years, existing works about POI recommendations exploit the separated side information such as geographical locations (e.g. longitude and latitude)~\cite{rahmani2019lglmf}, categories~\cite{lin2018cate}, etc. However, Side information is predominantly in the form of bipartite graphs. Applying this information to the representation learning can improve the performance of POI recommendation to some extent, but does not structure the information well. The information is isolated from each other in different bipartite graphs, which makes it difficult to propagate information between graphs.
In this section, we first introduce the conceptional graph for our UrbanKG then the idea of knowledge disentanglement, which aims to disentangle geographical and functional attributes.
\subsection{Urban Knowledge Graph Construction}

As mentioned before, more integrated, structured and fine-grained information needs to be obtained and applied to the recommendation tasks.
In this paper, we take the pioneering step\footnote{It is worth mentioning although there are several works~\cite{chen2021multi, zhang2019kean} trying to build the knowledge graph 
for POIs, these graphs are separated bipartite graphs and the knowledge is fragmented.} to build a large-scale Urban Knowledge Graph (UrbanKG) that contains rich entities and relations to provide information of POIs more structured, naturally and comprehensively.
Specifically, we first collect various POI information and cities' geographical data from Tencent Map, a famous Map service in China.
We then define a conceptual graph that reconstruct these data into the urban knowledge graph.
For the geographical attributes of POIs, we discard the confusing conventional features like longitude and latitude (specific locations), replacing them with entities like business areas and regions which exist in real cities and can be perceived by users. On the other hand, for the functional attributes of POIs, we collect three-level categories and brands, which serve as entities in UrbanKG and play important roles in fine representations of POIs.
The correlations between POI and these supplementary urban entities are depicted as different relations in UrbanKG. For example, \textit{LocateAt} builds the correlation between a POI and a region, showing that the POI locates at the given region. \textit{BrandOf} represents the correlation between a POI and a brand, indicating the brand of the given POI.

There are 7 types of entities including POIs, business areas, regions, brands and categories, and 16 types of relations.
We then collect the user check-in logs in WeChat, which is the largest instant messaging App in China. Through filtering the overlapped POIs, we build datasets of two largest cities of China, Beijing and Shanghai for UrbanKG-enhanced POI recommendation.
\begin{figure}
\centering
    \includegraphics[width=8.5cm]{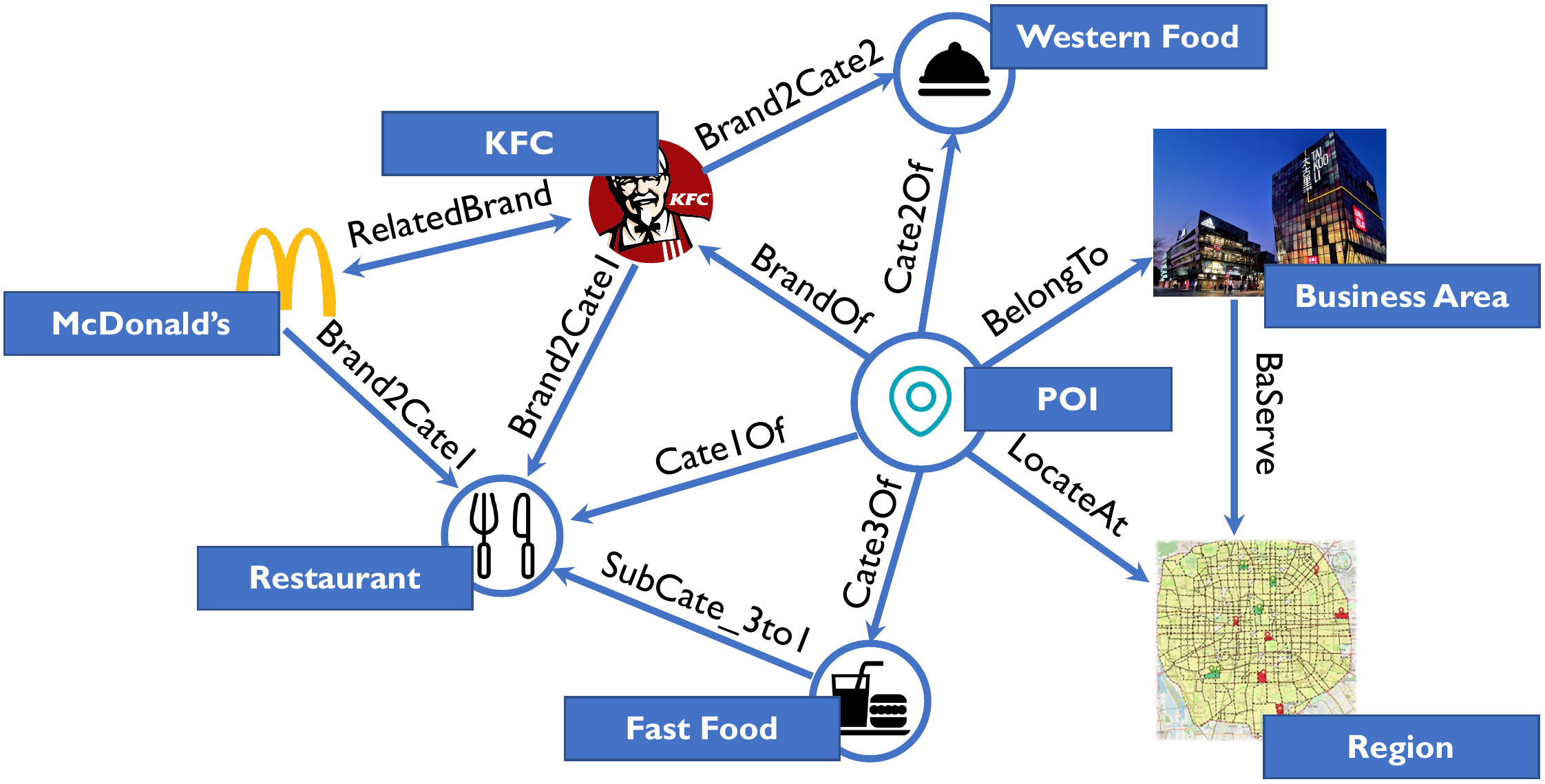}
    \caption{An example of part of multiple entities and relations in our UrbanKG. For the given POI, UrbanKG provides its geographical attributes like business area and region it locates at, and functional attributes such as its brand and three-level categories.}
        \vspace{-0.4cm}
    \label{fig:kgexample}
\end{figure}

Let $\mathcal{E}$ be the set of urban entities and $\mathcal{R}$ be the set of relations. UrbanKG can be represented as a collection of triplets $\mathcal{G} = \{(h,r,t)|h,t \in \mathcal{E}, r\in \mathcal{R}\}$, in which each triplet depicts the relation $r$ exists between the head entity $h$ and the tail entity $t$. For instance, \textit{(Apple Store East Nanjing Road, BrandOf, Apple)} indicates that the brand of POI \textit{Apple Store East Nanjing Road} is \textit{Apple}. To visualize the UrbanKG we build, we show an example of multiple entities and relations in our UrbanKG in Figure~\ref{fig:kgexample}.\footnote{The scale of our UrbanKG will be discussed in Section~\ref{sec::exp}.}

By constructing UrbanKG, we integrate knowledge into a structured graph rather than multiple bipartite graphs, breaking the isolation of information in previous works. Meanwhile, the UrbanKG structure broadens the way of information propagation and makes it accesible to exploit large-scale graph neural networks, which facilitates the representation learning.

\begin{table}[t]
    \centering
    \footnotesize
    \caption{Relations in the urban knowledge graph (``F" and ``G" refer to ``Functional" and ``Geographical", respectively.}
        \vspace{-0.4cm}
    \begin{tabular}{c|c|c}
        \hline
        \textbf{Relations} & \textbf{(\textit{head}, \textit{tail})} & \textbf{Type}\\
        \hline
        BaServe & (\textit{Business area}, \textit{Region})&G\\
        \hline
        BelongTo & (\textit{POI}, \textit{Business area})&G\\
        \hline
         BorderBy & (\textit{Region}, \textit{Region})&G\\
        \hline
         LocateAt & (\textit{POI}, \textit{Region})&G\\
        \hline
         NearBy & (\textit{Region}, \textit{Region})&G\\
         \hline
         Brand2Cate1 & (\textit{Brand}, \textit{Cate1})&F\\
        \hline
         Brand2Cate2 & (\textit{Brand}, \textit{Cate2})&F\\
        \hline
         Brand2Cate3 & (\textit{Brand}, \textit{Cate3})&F\\
        \hline
         BrandOf & (\textit{POI}, \textit{Brand})&F\\
        \hline
         Cate1Of & (\textit{POI}, \textit{Cate1})&F\\
        \hline
         Cate2Of  & (\textit{POI}, \textit{Cate2})&F\\
        \hline
         Cate3Of  & (\textit{POI}, \textit{Cate3})&F\\
        \hline
         RelatedBrand  & (\textit{Brand}, \textit{Brand})&F\\
        \hline
         SubCate\_2to1  & (\textit{Cate2}, \textit{Cate1})&F\\
        \hline
         SubCate\_3to1 & (\textit{Cate3}, \textit{Cate1})&F\\
        \hline
         SubCate\_3to2 & (\textit{Cate3}, \textit{Cate2})&F\\
         \hline
    \end{tabular}
    \vspace{-0.4cm}
    \label{tab:relations}
\end{table}
\subsection{Knowledge Disentanglement}
Different from traditional recommendation scenarios, such as e-commerce, where users can visit any item without effort, POI recommendations should highly consider the user's cost of visitation. Specifically, users tend to check-in POIs around their range of activity, showing the strong impact of geographical constraint. 
From the perspective of causal inference, geographical factor not only characterizes POIs but also affects the user-POI interaction, playing as a confounder.

Nevertheless, such confounder effect has not been explored by existing works.
Fortunately, our built UrbanKG provides a precious opportunity to consider it.
Specifically, UrbanKG mainly intergrates two kinds of POI information, geographical and functional.
Therefore, the relations are classified to two types, geographical type and funtional type based on the information they provide, as shown in Table~\ref{tab:relations}. For example, BelongTo is a geographical relation and it denotes the business area where a POI is affiliated. BrandOf and Cate1Of are functional relations and they depict the brand and the coarse-level category of POIs, respectively. The geographical-type relation is abbreviated as $G$, and the functional-type relation is abbreviated as $F$ in Table \ref{tab:relations}. Furthermore, we divide our constructed UrbanKG into two subgraphs, called the \textit{geographical graph} and the \textit{functional graph}, according to the type of relations. Geographical graph contains the relations of the geographical type and geographical entities like business areas and regions. Correspondingly, functional graph contains the relations of the functional type and functional entities such as brands, coarse-level, mid-level and fine-level categories of POIs. It should be noted that the same POIs are contained both in the \textit{geographical graph} and the \textit{functional graph}.

In the following, we will introduce our UKGC model learning from both user-POI interaction data and the UrbanKG for recommendation.

\begin{figure*}
\centering
    \includegraphics[height=5.3cm]{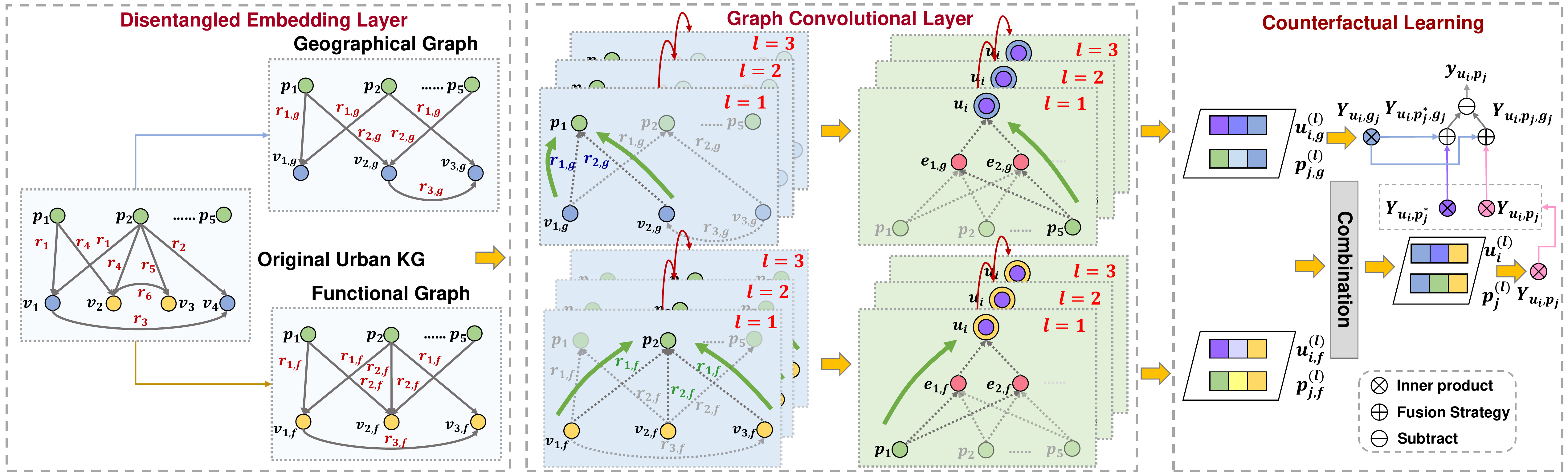}
    \caption{The framework of our proposed model UKGC. UKGC contains three parts: Disentangled Embedding Layer, Graph Convolutional Layer and Counterfactual Learning.}
        \vspace{-0.4cm}
    \label{fig:modelfig}
\end{figure*}

	\vspace{-0.4cm}
\section{Methodology} \label{sec::method}
To tackle the geographical bias of check-in behaviors and get a better POI recommendation performance, we propose an effective model called UKGC, which is illustrated in Figure~\ref{fig:modelfig}. It consists of three parts:
\begin{itemize}[leftmargin=*,partopsep=0pt,topsep=0pt]
\setlength{\itemsep}{0pt}
\setlength{\parsep}{0pt}
\setlength{\parskip}{0pt}
    \item \textbf{Disentangled Embedding Layer:} We propose to utilize two sets of embeddings for geographical and functional attributes of POIs and user interests, to make these two aspect of attributes disentangled.
    \item \textbf{Graph Convolutional Layer:} In order to learn disentangled representations of geographical and functional attributes, we train two sets of embeddings with geographical graph and functional graph derived from UrbanKG and user-POI interaction graph by information propagation via graph convolutional network.
    \item \textbf{Counterfactual Learning:} Finally, we present the causal graph where geographical attributes directly affect on the probability of interaction and introduce counterfactual inference to alleviate the geographical bias and for more accurate POI recommendations.
\end{itemize}
\subsection{Disentangled Embedding Layer}
To learn fine-grained embeddings represented both in geographical and functional aspects, we assign two chunks of embeddings for users and POIs, corresponding to the knowledge disentanglement in Section~\ref{sec::UKG}. 
The geographical and functional embeddings can be denoted by $g$ and $f$ in the subscript, respectively. 

We define the geographical embedding and the functional embedding of user $u_i$ (POI $p_j$) as $\mathbf{u}_{i,g}, \mathbf{u}_{i,f}\in \mathbb{R}^d$ ($\mathbf{p}_{j,g}, \mathbf{p}_{j,f} \in \mathbb{R}^d$), where $d$ denotes the embedding size. And we describe the non-POI entity $v_k$ in the geographical graph with an embedding vector $\mathbf{v}_{k,g} \in \mathbb{R}^d$, and $v_h$ in the functional graph with $\mathbf{v}_{h,f} \in \mathbb{R}^d$. Then the parameter embedding matrix of the geographical graph and the functional graph can be denoted as: 
\begin{equation}\label{fml::emb}
\small
\begin{aligned}
    \mathbf{E}_g = [\mathbf{u}_{1,g}; \cdots; \mathbf{u}_{N,g}; \mathbf{p}_{1,g}; \cdots; \mathbf{p}_{M,g}; \mathbf{v}_{1,g}; \cdots; \mathbf{v}_{W,g}],\\
    \mathbf{E}_f = [\mathbf{u}_{1,f}; \cdots; \mathbf{u}_{N,f}; \mathbf{p}_{1,f}; \cdots; \mathbf{p}_{M,f}; \mathbf{v}_{1,f}; \cdots; \mathbf{v}_{Q,f}].
\end{aligned}
\end{equation}
where $N$ and $M$ refer to the number of users and POIs, respectively. $W$ and $Q$ refer to the number of non-POI entities in the geographical graph and the functional graph.

To connect the user-POI interaction graph with the knowledge graph,
Wang et al~\cite{wang2021kgin} proposed that the user-POI interactions are driven by several~\textit{intents}, which can be represented by the distribution of relations in KG.
Inspired by it, here we model users' two sets of intentions, geographical and functional intents simultaneously, denoted as $\mathcal{I}_{g}$ and $\mathcal{I}_{f}$.

The embedding of the $i$-th intent in the geographical graph is defined as follows,

\begin{equation} \label{fml::int}
\small
\mathbf{e}_{i,g} = \sum_{{r_{i,g}}\in \mathcal{R}_{g}}\alpha(i,j)\mathbf{r}_{j,g},
\end{equation}
where $\mathcal{R}_{g}$ represents the set of relations in the geographical graph, and $\mathbf{r}_{j,g}$ is the embedding of the $j$-th relation in the geographical graph. $\alpha(i,j)$ denotes the attention score, which can be formulated as follows,
\begin{equation} \label{fml::alpha}
\small
\alpha(i,j) = \frac{\exp{(s_{ij})}}{\sum_{\mathbf{r}_{k,g}\in \mathcal{R}_{g}}\exp{(s_{ik})}},
\end{equation}
where $s_{ij}$ is the trainable weight for the $j$-th relation and the $i$-th intent, which is applicable to all users. 
The embedding of the $i$-th intent in the functional graph, i.e. $\mathbf{e}_{i, f}$, is derived similarly.

Following the previous work~\cite{Sz_kely2007,Sz_kely2009}, we utilize the distance correlation as the regularizer to make sure user intents are independent to carry as much information as possible. As follows, 
\begin{equation}\label{fml::ind}
\small
    \mathcal{L}_{\mathrm{IND}} = \sum_{\mathbf{e}_i, \mathbf{e}_j \in \mathcal{I}_{g}, i \neq j}dCor(\mathbf{e}_i, \mathbf{e}_{j}),
\end{equation}
where $dCor(\cdot)$ represents the distance correlation between two intents, formulated as follows,
\begin{equation}\label{fml::dcor}
\small
    dCor(\mathbf{e}_i, \mathbf{e}_{j}) = \frac{dCov(\mathbf{e}_i, \mathbf{e}_{j})}{\sqrt{dVar(\mathbf{e}_i)\cdot dVar(\mathbf{e}_j)}},
\end{equation}
where $dCov(\cdot)$ indicates the distance covariance of two embeddings, and $dVar(\cdot)$ indicates the distance variance of each embedding. 
\subsection{Graph Convolutional Layer}
User embeddings should contain information from both KG and user-POI interaction.
Therefore, we propose the embedding propagation layer as follows,
\begin{equation}\label{fml::prop}
\small
\mathbf{u}_{i,g}^{(1)} =\mathbf{u}_{i,g}^{(0)}+ \frac{1}{|\mathcal{N}_{u_{i}}||\mathcal{I}_g|}\sum_{e_{j,g} \in \mathcal{I}_g}\sum_{{p}_k \in \mathcal{N}_{{u}_{i}}}\beta(i,j)\mathbf{e}_{j,g}\odot \mathbf{p}_{k,g}^{(0)},
\end{equation}
where $\mathbf{p}_{k,g}^{(0)}$ and $\mathbf{u}_{i,g}^{(0)}$ represent the embedding of the $k$-th POI and the $i$-th user respectively in the geographical graph, defined in Equation~\ref{fml::emb}. \(\odot\) is the Hadamard product. \(\mathcal{N}_{{u}_{i}} = \{{p}_k| (u_i, p_k) \in \mathcal{O^+}\}\) represents the positive POI sample related to user $u_i$, and $\beta(i,j)$ is the attention score of intent $e_{j,g}$ about user $u_i$, formulated as:
\begin{equation}\label{fml::beta}
\small
\beta(i,j)=\frac{\exp{(\mathbf{e}_{j,g}^{T}\mathbf{u}_{i,g}^{(0)})}}{\sum_{e_{t,g}\in  \mathcal{I}_g}\exp{(\mathbf{e}_{t,g}^{T}\mathbf{u}_{i,g}^{(0)})}},
\end{equation}
where $\mathbf{u}_{i,g}^{(0)}$ is the embedding of the $i$-th user in the geographical graph, defined in Equation~\ref{fml::emb}. $\mathbf{e}_{j,g}$ is the embedding of the $j$-th intent of the geographical graph.

For the embeddings of POIs in the geographical graph, we derive them by similar propagation rules. The difference is that the embeddings of POIs are obtained through the information propagation on a partitioned urban knowledge graph, based on various relations. POIs have no preference for information from a certain relation, so we do not need to assign the attention score for POI $p_i$ and relation $r_j$. Similarly, the embeddings of non-POI entities $\mathbf{V}_g = [\mathbf{v}_{1,g}; \cdots;  \mathbf{v}_{W,g}]$ in the geographical graph can also be acquired analogously, propagating according to the same manner.

After captured the first-order connectivity of nodes in the graph, we extend more layers to better acquire information from high-order neighbors. Propagated for \(l\) layers, we can obtain embeddings recursively:
\begin{equation}\label{fml::players}
\small
\mathbf{p}_{i,g}^{(l)} = \mathbf{p}_{i,g}^{(l-1)}+\frac{1}{|\mathcal{N}_{p_{i,g}}|}\sum_{(r_{j,g},v_{k,g}) \in \mathcal{N}_{p_{i,g}}}\mathbf{r}_{j,g}\odot \mathbf{v}_{k,g}^{(l-1)}  
\end{equation}

\begin{equation}\label{fml::users}
\small
\mathbf{u}_{i,g}^{(l)} =\mathbf{u}_{i,g}^{(l-1)}+ \frac{1}{|\mathcal{N}_{u_{i}}||\mathcal{I}_g|}\sum_{e_{j,g} \in \mathcal{I}_g}\sum_{p_k\in \mathcal{N}_{u_{i}}}\beta(i,j)\mathbf{e}_{j,g}\odot \mathbf{p}_{k,g}^{(l-1)}
\end{equation}
\begin{equation}
\small
\mathbf{p}_{i,f}^{(l)} = \mathbf{p}_{i,f}^{(l-1)}+\frac{1}{|\mathcal{N}_{p_{i,f}}|}\sum_{(r_{t,f},v_{h,f}) \in \mathcal{N}_{p_{i,f}}}\mathbf{r}_{t,f}\odot \mathbf{v}_{h,f}^{(l-1)}
\end{equation}
\begin{equation}
\small
\mathbf{u}_{i,f}^{(l)} =\mathbf{u}_{i,f}^{(l-1)}+ \frac{1}{|\mathcal{N}_{u_{i}}||\mathcal{I}_f|}\sum_{e_{t,f} \in \mathcal{I}_f}\sum_{p_h\in \mathcal{N}_{u_{i}}}\beta(i,t)\mathbf{e}_{t,f}\odot \mathbf{p}_{h,f}^{(l-1)}
\end{equation}

Inspired by the information combination method proposed by the previous work~\cite{wei2019MMGCN}, we integrate two sets of embeddings from two partitioned knowledge graphs via the linear combination and get final embeddings of the $i$-th user and the $i$-th POI, formulated as follows,
\begin{equation}\label{fml::uulti}
\small
\mathbf{u}_i=\frac{1}{2}(\mathbf{u}_{i,g}^{(l)}+\mathbf{u}_{i,f}^{(l)}), \mathbf{p}_i=\frac{1}{2}(\mathbf{p}_{i,g}^{(l)}+\mathbf{p}_{i,f}^{(l)}).
\end{equation}

\subsection{Counterfactual Learning}
As mentioned in Section~\ref{sec::intro}, the geographical factor can be considered as the confounder from the causal inference view~\cite{pearl2018book}. To address it, here we first present the causal graph of the recommendation task and then propose a counterfactual learning method.

\subsubsection{Causal Graph in POI Recommendation} 
\begin{figure}[t]
    \centering
    \subfigure[]{
    \label{causalgraph}
    \includegraphics[width=0.25\linewidth]{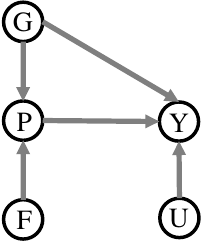}}
    \hspace{14mm}
    \subfigure[]{
    \label{realworld}
    \includegraphics[width=0.25\linewidth]{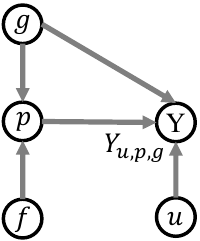}}
    \subfigure[]{
    \label{counterfactual}
    \includegraphics[width=0.29\linewidth]{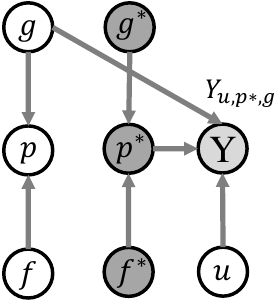}
    }
    \hspace{10mm}
    \subfigure[]{
    \label{reference}
    \includegraphics[width=0.25\linewidth]{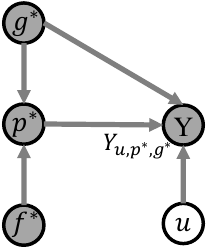}
    }    \vspace{-0.4cm}
    \caption{The causal graphs for the factual world, the counterfactual world and the reference status, where * denotes the reference value of random variables, (a) denotes the proposed causal graph, (b) denotes the factual world, (c) denotes the counterfactual world, (d) denotes the reference status.}
        \vspace{-0.4cm}
    \label{fig::causal}
\end{figure}
The causal graph is in the form of directed acyclic graph $\mathcal{G} = \{\mathcal{N}, \mathcal{E}\}$, where $\mathcal{N}$ denotes the variables and $\mathcal{E}$ denotes the causal relations. The capital letters in the causal graph are random variables, and the lowercase letters are corresponding values of them. The directed edge indicates that the head node has a causal effect to the tail node. As shown in the Figure~\ref{causalgraph}, $\mathrm{P}$ represents POIs, $\mathrm{G}$ and $\mathrm{F}$ represents the geographical and functional attributes of the POI, respectively. $\mathrm{U}$ represents users, and $\mathrm{Y}$ means the probability of the user-POI interaction.
In the conventional POI recommendation models, POIs contain geographical and functional attributes, reflected in the causal graph as $\mathrm{G} \rightarrow \mathrm{P}$ and $\mathrm{F} \rightarrow \mathrm{P}$. As previously stated, there is a causal relation $\mathrm{G} \rightarrow \mathrm{Y}$, to model the influence of the direct effect of geographical attributes, i.e., geographical bias. 
According to the causal graph, the value of $\mathrm{Y}$ and $\mathrm{P}$ can be derived as
\begin{equation}\label{fml::causal}
\small
\begin{split}
    \mathrm{Y}_{u_i,p_j,g_j} &= \mathrm{Y}(\mathrm{U}=u_i, \mathrm{P}=p_j, \mathrm{G}=g_j),\\
    p_j &= \mathrm{P}(\mathrm{G}=g_j, \mathrm{F}=f_j),
\end{split}
\end{equation}
where $\mathrm{Y}(\cdot)$ and $\mathrm{P}(\cdot)$ denotes the function to get the value of $\mathrm{Y}$ and $\mathrm{P}$.
\subsubsection{Counterfactual Learning}

For the $i$-th user $u_i$, the \textit{total effect}~(TE) of the $j$-th POI's attributes $\mathrm{G}=g_j$ and $\mathrm{F}=f_j$ on the likelihood of interaction can be formulated as follows,
\begin{equation}\label{fml::TE}
\small
\mathrm{TE} = \mathrm{Y}_{u_i,p_j,g_j} - \mathrm{Y}_{u_i,p^*_j,g^*_j},
\end{equation}
where $p^*_j=\mathrm{P}(\mathrm{G}=g^*_j, \mathrm{F}=f^*_j)$ and $g^*_j$ are the reference value of $\mathrm{G}$ and $\mathrm{P}$, respectively. \textit{Reference value} means that the POI does not have its own characteristics, and the corresponding attributes ($\mathrm{G}$ or $\mathrm{P}$) of POIs are all the same, taking the average value. 

The geographical bias can be expressed as the \textit{natural direct effect}~(NDE) of geographical attributes on the probability of interaction, which can be formulated as follows,
\begin{equation}\label{fml::nde}
\small
\mathrm{NDE} = \mathrm{Y}_{u_i,p^{*}_j,g_j}-\mathrm{Y}_{u_i,p^{*}_j,g^{*}_j},
\end{equation}
where $p^*_j$, $g^*_j$ and $f^*_j$ are the reference value of $\mathrm{P}$, $\mathrm{G}$ and $\mathrm{F}$, respectively. $\mathrm{Y}_{u_i,p^*_j,g_j}$ refers to the counterfactual world, representing \textit{what the interaction probability would be if a user can only obtain the geographical information of the POI.}

To eliminate the geographical bias, we need to subtract the $\mathrm{NDE}$ of $\mathrm{G}$ from $\mathrm{TE}$. The implication is that users will not interact based on geographical information alone, but make decisions taking a combination of geographical and functional information of the POI into consideration. 
Because there is no direct effect path $\mathrm{F} \rightarrow \mathrm{Y}$, we do not need to subtract the $\mathrm{NDE}$ of $\mathrm{F}$ and then get the \textit{total indirect effect}~(TIE) of geographical attributes $\mathrm{G}$ and functional attributes $\mathrm{F}$ after subtracting the $\mathrm{NDE}$ of $\mathrm{G}$. Now, we use the $\mathrm{TIE}$ as the debiased interaction prediction score between user $u_i$ and POI $p_j$, formulated as follows,
\begin{equation}\label{fml::tie}
\small
 \hat{y}_{u_i,p_j}=\mathrm{TIE} = \mathrm{TE} - \mathrm{NDE} = \mathrm{Y}_{u_i,p_j,g_j}-\mathrm{Y}_{u_i,p^*_j,g_j}.
\end{equation}

We split $\mathrm{Y}_{u_i,p_j,g_j}$ and $\mathrm{Y}_{u_i, p^*_j, g_j}$ into three kinds of scores, $\mathrm{Y}_{u_i,p_j}$, $\mathrm{Y}_{u_i,g_j}$ and $\mathrm{Y}_{u_i, p^*_j}$, where $g_j$ represents the geographic attributes of POI $p_j$. Inspired by previous work~\cite{cadene2019rubi,niu2020counterfactual,wang2021clicks}, we calculate the values of these two components adopting the \textit{fusion strategy}, formulated as follows,
\begin{equation}\label{fml::te}
\small
\begin{aligned}
    \mathrm{Y}_{u_i,p_j,g_j} = f(\mathrm{Y}_{u_i,p_j}, \mathrm{Y}_{u_i,g_j}), \\
    \mathrm{Y}_{u_i,p^*_j,g_j} = f(\mathrm{Y}_{u_i, p^*_j}, \mathrm{Y}_{u_i,g_j}),
    \end{aligned}
\end{equation}
where $f(\cdot)$ represents the fusion function. 
One promising choice is as follows,
\begin{equation}\label{fml::mt}
\small
    f(\mathrm{Y}_{u_i,p_j}, \mathrm{Y}_{u_i,g_j}) = \mathrm{Y}_{u_i,p_j} * \tanh{(\mathrm{Y}_{u_i,g_j})}.
\end{equation}

Each components stated before can be calculated by the inner product of embeddings, which refer to the ultimate embeddings after propagation in (\ref{fml::uulti}), formulated as follows,
\begin{equation}\label{fml::yup}
\small
    \mathrm{Y}_{u_i,p_j} = \mathbf{u}_{i}^{T}\mathbf{p}_{j}, \mathrm{Y}_{u_i,g_j}=\mathbf{u}_{i,g}^{T}\mathbf{p}_{j,g},
\end{equation}
\begin{equation}\label{fml::pstar}
\small
    \mathrm{Y}_{u_i, p^*_j}= \mathbb{E}(\mathrm{Y}_{u_i,P})=\frac{1}{|\mathcal{P}|}\sum_{p_t \in \mathcal{P}}\mathrm{Y}_{u_i,p_t}
\end{equation}
where $\mathcal{P}$ denotes the set of POIs and $|P|$ is the cardinality of the set $|P|$.
\subsection{Model Optimization} 
To optimize the prediction of $\mathrm{Y}_{u_i,p_j,g_j}$ in the factual world, we utilize the BPR loss~\cite{rendle2009bpr} to distinguish between positive and negative samples as follows,
\begin{equation}\label{fml::bprloss}
\small
    \mathcal{L}_{F}=\sum_{(u_i,p_j,p_k) \in \mathcal{O}} -\ln{\sigma(\mathrm{Y}_{u_i,p_j}-\mathrm{Y}_{u_i,p_k})}
\end{equation}
where $\mathcal{O}=\{(u_i,p_j,p_k)|(u_i,p_j)\in \mathcal{O^+},(u_i,p_k)\in \mathcal{O^-}\}$ denotes the training set, $\mathcal{O^+}$ represents positive interactions, i.e. observed check-in data between users and POIs, and $\mathcal{O^-}$ represents negative interactions, i.e. unobserved. $\sigma(\cdot)$ denotes the sigmoid function.

We also need to better learn the prediction of $\mathrm{Y}_{u_i,p^*_j,g_j}$ in the counterfactual world. Note that for user $u_i$, $\mathrm{Y}_{u_i, p^*_j} = \mathbb{E}(\mathrm{Y}_{u_i,P})$ is identical for every POI. Hence, we only need to use the BPR loss to distinguish $\mathrm{Y}_{u_i,g_j}$ and $\mathrm{Y}_{u_i,g_k}$ as follows,
\begin{equation}\label{fml::bprlossgeo}
\small
    \mathcal{L}_{C}=\sum_{(u_i,g_j,g_k) \in \mathcal{O}} -\ln{\sigma(\mathrm{Y}_{u_i,g_j}-\mathrm{Y}_{u_i,g_k})}
\end{equation}
where the definition of $\mathcal{O}$ is similar to (\ref{fml::bprloss}), and $g_j$ and $g_k$ are geographical attributes of $p_j$ and $p_k$, respectively. 

By combining $\mathcal{L}_{F}$, $\mathcal{L}_{C}$, and independent loss $\mathcal{L}_{\mathrm{IND}}$ in (\ref{fml::ind}), the full objective function can be formulated as follows,
\begin{equation}\label{fml::totalloss}
\small
\begin{split}
    \mathcal{L} = \mathcal{L}_{F} + \lambda_1(\mathcal{L}_{\mathrm{IND}g} + \mathcal{L}_{\mathrm{IND}f}) + \lambda_2||\Theta||_2\\
    + \alpha(\mathcal{L}_{C} + \lambda_2||\Theta_{g}||_2),
\end{split}
\end{equation}
where $\lambda_1$ and $\lambda_2$ represents the hyperparameters controlling the independence loss and $L_2$ regularization term, respectively. $\Theta$ denotes trainable parameters in the entire model. Trainable parameters in the geographical graph $\Theta_{g}$ are also there to prevent the overfitting of $\mathrm{Y}_{u_i,g_j}$
. $\alpha$ is the hyperparameter controlling the relative weight of learning in the factual world compared with the counterfactual world.
	\section{Experiments}\label{sec::exp}
In this section, we conduct experiments to show the effectiveness of the proposed UKGC model and we aim to answer the following research questions:
\begin{itemize}[leftmargin=*,partopsep=0pt,topsep=0pt]
\setlength{\itemsep}{0pt}
\setlength{\parsep}{0pt}
\setlength{\parskip}{0pt}
    \item \textbf{RQ1: } How does our proposed UKGC model perform compared with different kinds of state-of-the-art recommendation methods?
    \item \textbf{RQ2: } How do different components in the proposed UKGC affect the model?
    \item \textbf{RQ3: } What are the influences of different settings of hyperparameters in our model?
    \item \textbf{RQ4: } Can the proposed UKGC model really capture geographical attributes and functional attributes of POIs well?
\end{itemize}
\begin{table*}[htbp]
    \centering
      \footnotesize
          \vspace{-0.4cm}
    \caption{Overall performance comparison.}
        \vspace{-0.4cm}
    \begin{tabular}{c|c|c|c|c|c|c|c|c}
       \hline
       \multicolumn{9}{c}{\textbf{Beijing}}\\
       \hline
       \textbf{Category}&\textbf{Model}&\textbf{AUC}&\textbf{Recall@20}&\textbf{Recall@40}&\textbf{Recall@60}&\textbf{NDCG@20}&\textbf{NDCG@40}&\textbf{NDCG@60}\\
       \hline
       CF-based&\textbf{LightGCN}&0.7867&\underline{0.0508}&\underline{0.0781}&\underline{0.0975}&\underline{0.0404}&\underline{0.0513}&\underline{0.0583}\\
       \hline
       \multirow{5}{*}{KG-based}&\textbf{CKE}&0.8243&0.0485&0.0753&0.0952&0.0386&0.0494&0.0565\\
      \cline{2-9}
       &\textbf{CFKG}&0.6558&0.0381&0.0577&0.0716&0.0312&0.0391&0.0440\\
    \cline{2-9}
       &\textbf{KGAT}&0.8159&0.0343&0.0569&0.0743&0.0257&0.0347&0.0409\\
      \cline{2-9}
       &\textbf{KGIN}&\underline{0.8488}&0.0475&0.0742&0.0940&0.0389&0.0496&0.0567\\
      \cline{2-9}
       &\textbf{UKGC}&\textbf{0.8824}&\textbf{0.0585}&\textbf{0.0874}&\textbf{0.1067}&\textbf{0.0507}&\textbf{0.0624}&\textbf{0.0693}\\
       \hline
             \hline
      \multicolumn{9}{c}{\textbf{Shanghai}}\\
       \hline
       \textbf{Category}&\textbf{Model}&\textbf{AUC}&\textbf{Recall@20}&\textbf{Recall@40}&\textbf{Recall@60}&\textbf{NDCG@20}&\textbf{NDCG@40}&\textbf{NDCG@60}\\
       \hline
       CF-based&\textbf{LightGCN}&0.7634&0.0468&0.0700&0.0893&0.0359&0.0445&0.0508\\
       \hline
       \multirow{5}{*}{KG-based}&\textbf{CKE}&0.8053&0.0377&0.0599&0.0762&0.0276&0.0359&0.0412\\
      \cline{2-9}
       &\textbf{CFKG}&0.6104&0.0255&0.0419&0.0527&0.0174&0.0235&0.0270\\
    \cline{2-9}
      &\textbf{KGAT}&0.8000&0.0301&0.0490&0.0645&0.0219&0.0289&0.0340\\
      \cline{2-9}
       &\textbf{KGIN}&\underline{0.8451}&\underline{0.0504}&\underline{0.0756}&\underline{0.0948}&\underline{0.0391}&\underline{0.0485}&\underline{0.0548}\\
      \cline{2-9}
       &\textbf{UKGC}&\textbf{0.8776}&\textbf{0.0543}&\textbf{0.0800}&\textbf{0.0997}&\textbf{0.0436}&\textbf{0.0532}&\textbf{0.0596}\\
       \hline
    \end{tabular}
    \vspace{-0.4cm}
    \label{tab:bjcompare}
\end{table*}

\begin{table}[t]
    \centering
    \footnotesize
    \label{tab:resauc}
    \caption{AUC performance comparison in two datasets.}
    \vspace{-0.4cm}
    \begin{tabular}{c|c|c|c}
    \hline
    \multicolumn{2}{c|}{}&\textbf{Beijing}&\textbf{Shanghai}\\
    \hline
    \textbf{Category}&\textbf{Model}&\textbf{AUC}&\textbf{AUC}\\
    \hline
    \multirow{3}{*}{Feature-based}&\textbf{FM}&0.8197&0.8026\\
    \cline{2-4}
    &\textbf{DeepFM}&0.8188&0.8090\\
    \cline{2-4}
    &\textbf{AutoInt}&0.8206&0.8277\\
    \hline
    CF-based&\textbf{LightGCN}&0.7867&0.7634\\
    \hline
    \multirow{5}{*}{KG-based}&\textbf{CKE}&0.8243&0.8053\\
    \cline{2-4}
    &\textbf{CFKG}&0.6558&0.6104\\
    \cline{2-4}
    &\textbf{KGAT}&0.8159&0.8000\\
    \cline{2-4}
    &\textbf{KGIN}&\underline{0.8488}&\underline{0.8451}\\
    \cline{2-4}
    &\textbf{UKGC}&\textbf{0.8824}&\textbf{0.8776}\\
    \hline
    \end{tabular}
\end{table}
\subsection{Experimental Settings}
\textbf{Datasets. }We construct UrbanKG datasets on the two most biggest cities in China, Beijing and Shanghai. We have also collect users' check-in data of which the POIs exists in the UrbanKG.
The statistics are shown in Table \ref{tab:statistic}. 
\begin{table}[t]
  \footnotesize
    \centering
        \vspace{-0.3cm}
    \caption{Statistics of two datasets.}
        \vspace{-0.3cm}
    \begin{tabular}{c|c|c|c}
    \hline
    \multicolumn{2}{c|}{}&\textbf{Beijing}&\textbf{Shanghai}\\
    \hline
    \hline
    \multirow{3}{*}{Check-in Data} & \#Users & $10,000$ & $10,000$\\
    &\#POIs&$177,602$&$169,123$\\
    &\#Check-in&$763,073$&$618,197$\\
    \hline
    \hline
    \multirow{3}{*}{Urban KG}&\#Entities&$181,817$&$173,504$\\
    &\#Relations&$20$&$20$\\
    &\#Triplets&$336,178$&$384,702$\\
    \hline
    \end{tabular}
    \vspace{-0.4cm}
    \label{tab:statistic}
\end{table}
\textbf{Baselines.} Although our work targets POI recommendation, it is not appliable to compare our model UKGC with other conventional GNN-based POI recommendation models. Because of our pineering step to build the UrbanKG, our datasets no longer contain the specific location information such as the longitude and latitude of POIs, which are widely used in existing POI recommendation models and collected in most user-POI interaction datasets. Besides, conventional GNN-based POI recommendation models~\cite{Lim2020STP,chang2020GPR} cannot make use of other information such as brands and categories provided in the UrbanKG, and then it may lead to unfairness to some extent. After considering the reasons above, we compared our model with three kinds of SOTA recommendation methods in conventional tasks, including feature-based methods\footnote{These methods are also known as Click-Through Rate (CTR) prediction models, of which AutoInt~\cite{song2019autoint} is the SOTA one.} (FM, DeepFM, and AutoInt), collaborative filtering (LightGCN), and KG-based methods (CFKG, CKE, KGAT, and KGIN). 
\textbf{Evaluation Metrics.} As for top-$K$ recommendation performance, we adopt a full ranking strategy to test the performance. 
We consider the interacted POIs to be the positive samples for every user and treat other POIs as negative. 
We use two widely-accepted ranking metrics, Recall@$K$ and NDCG@$K$. 
However, the full-ranking task for traditional feature-based methods
is not acceptable due to the extremely high computation cost.
Following the original papers~\cite{guo2017deepfm,song2019autoint}, we use the AUC metric to evaluate these models with sampled negative items.\\
\textbf{Hyper-Parameter Settings.} We implement our UKGC model in PyTorch.
We fix embedding size as 64, which is commonly accepted in existing works~\cite{zhang2018cfkg, wang2019kgat, wang2021kgin}. Since we have deployed two parts of disentangled embeddings, for each part we assign the size as 32 for fair comparison. 
We use the Adam optimizer, set the batch size as 1024 and use Xavier initilizations for model parameters. 
We adopt a careful grid search of learning rate in $\{10^{-4},10^{-3},10^{-2}\}$ and the coefficient of $L_2$ normalization in $\{10^{-5},10^{-4},\cdots,10^{-1}\}$. 
For CFKG, CKE, KGAT and KGIN, we tune the message dropout ratio and node dropout radio in $\{0.0,0.1,\cdots,0.8\}$. 
For the multi-task learning weight $\alpha$, we search it in $\{0,0.25,0.5,0.75,1,2,3,4,5\}$. 
Moreover, early stopping is used, where we stop training if the validation performance does not increase for ten successive epochs.
\subsection{Overall Performance (RQ1)} 
We show the performance of all models in Table~\ref{tab:bjcompare}.
It should be noted that these experimental results are averages obtained after multiple times of experiments with different random seeds. 
From the experimental data, we can obtain the following results.
\begin{itemize}[leftmargin=*,partopsep=0pt,topsep=0pt]
\setlength{\itemsep}{0pt}
\setlength{\parsep}{0pt}
\setlength{\parskip}{0pt}
    \item \textbf{Our proposed UKGC method steadily achieves 
    the best performance.
    } Our model improves over the best baseline \textit{w.r.t.} AUC by $3.80\%$, Recall by $13.51\%-23.16\%$, and NDCG by $22.22\%-30.33\%$ in Beijing dataset, and improves over \textit{w.r.t.} AUC by $3.85\%$, Recall by $5.16\%-7.74\%$, and NDCG by $8.75\%-11.51\%$ in Shanghai dataset. 
    Our UKGC method achieves significant improvements on all metrics.
    \item \textbf{Necessity and effectiveness of building UrbanKG.}
    Compared with feature-based methods, our model significantly improves \textit{w.r.t.} AUC by $7.65\%$ in Beijing dataset, $5.98\%$ in Shanghai dataset\footnote{A 0.01-level of AUC improvement can be claimed as significant~\cite{song2019autoint,guo2017deepfm}}.
    It indicates the necessity and effectiveness of building and utilizing structural knowledge rather than separated features for the POI recommendation.
    We can also observe KGIN which is the SOTA KG-based recommendation model, serves as the best baseline, with the second best performance on Shanghai, only under our UKGC, and competitive performance on Beijing. 
    It further demonstrates that our built UrbanKG can help improve the recommendation performance.

    \item \textbf{It is essential to leverage UrbanKG appropriately.} 
    We can observe that KG-based methods, CKE, CFKG and KGAT, get poorer performance than feature-based methods \textit{w.r.t.} AUC in Shanghai dataset, which reveals that they cannot exploit urban knowledge appropriately.  
    They also get poorer performance than LightGCN \textit{w.r.t.} Top-K metrics. 
    Besides, KGIN either cannot beat LightGCN on Top-K metrics for the Beijing dataset.
    These observations reveal that special designs are required to leverage UrbanKG well. In our method, we design disentangled embeddings along with corresponding GCN layers and counterfactual learning, fully modeling geographical and functional attributes, thus achieves the best performance. 
\end{itemize}
\subsection{Ablation Study (RQ2)}
\begin{figure}[t]
    \centering
    \vspace{-0.2cm}
    \subfigure[Recall.]{
    \label{abrecallbj}
    \includegraphics[width=0.47\linewidth]{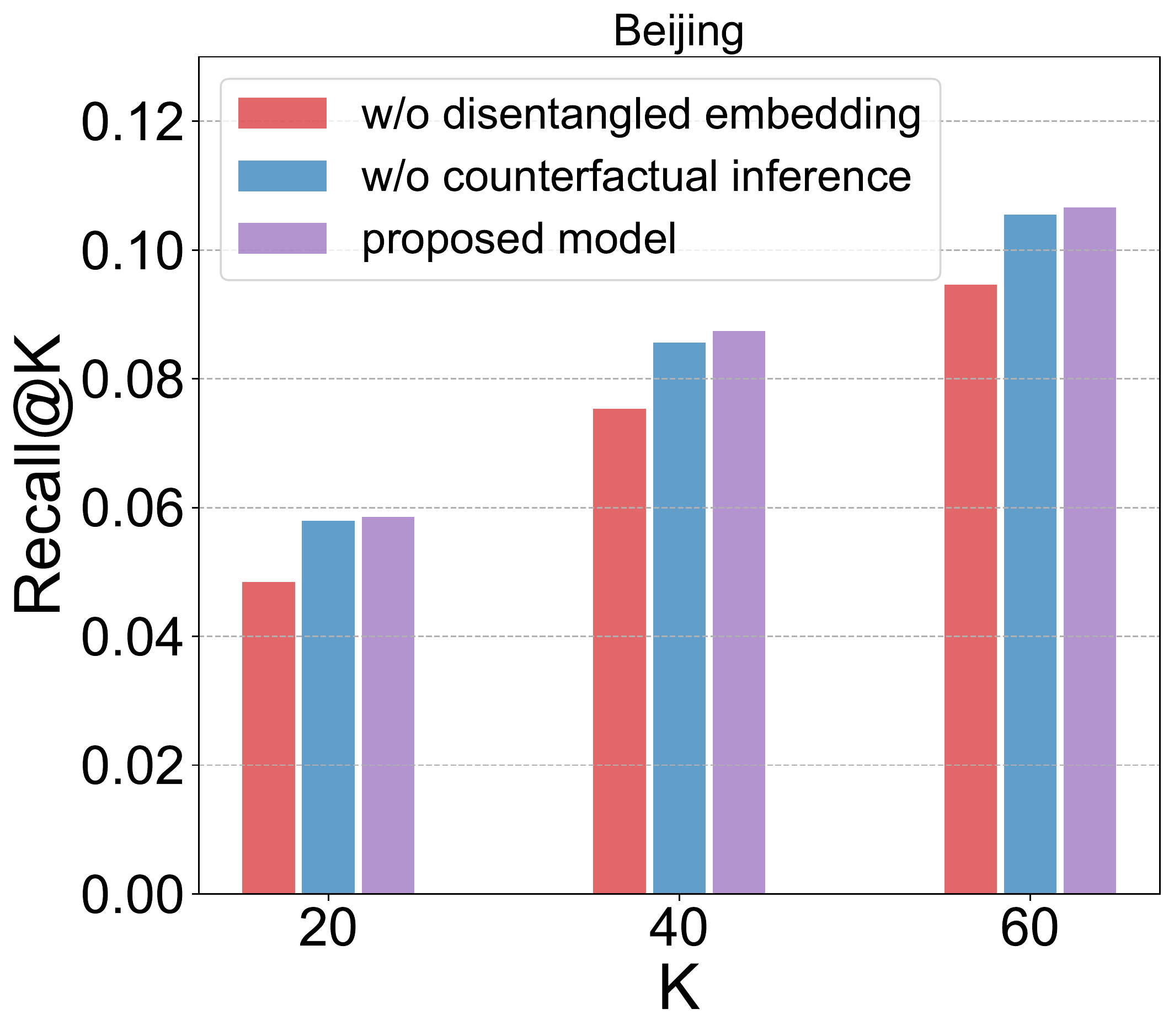}}\subfigure[NDCG.]{
    \label{abndcgbj}
    \includegraphics[width=0.47\linewidth]{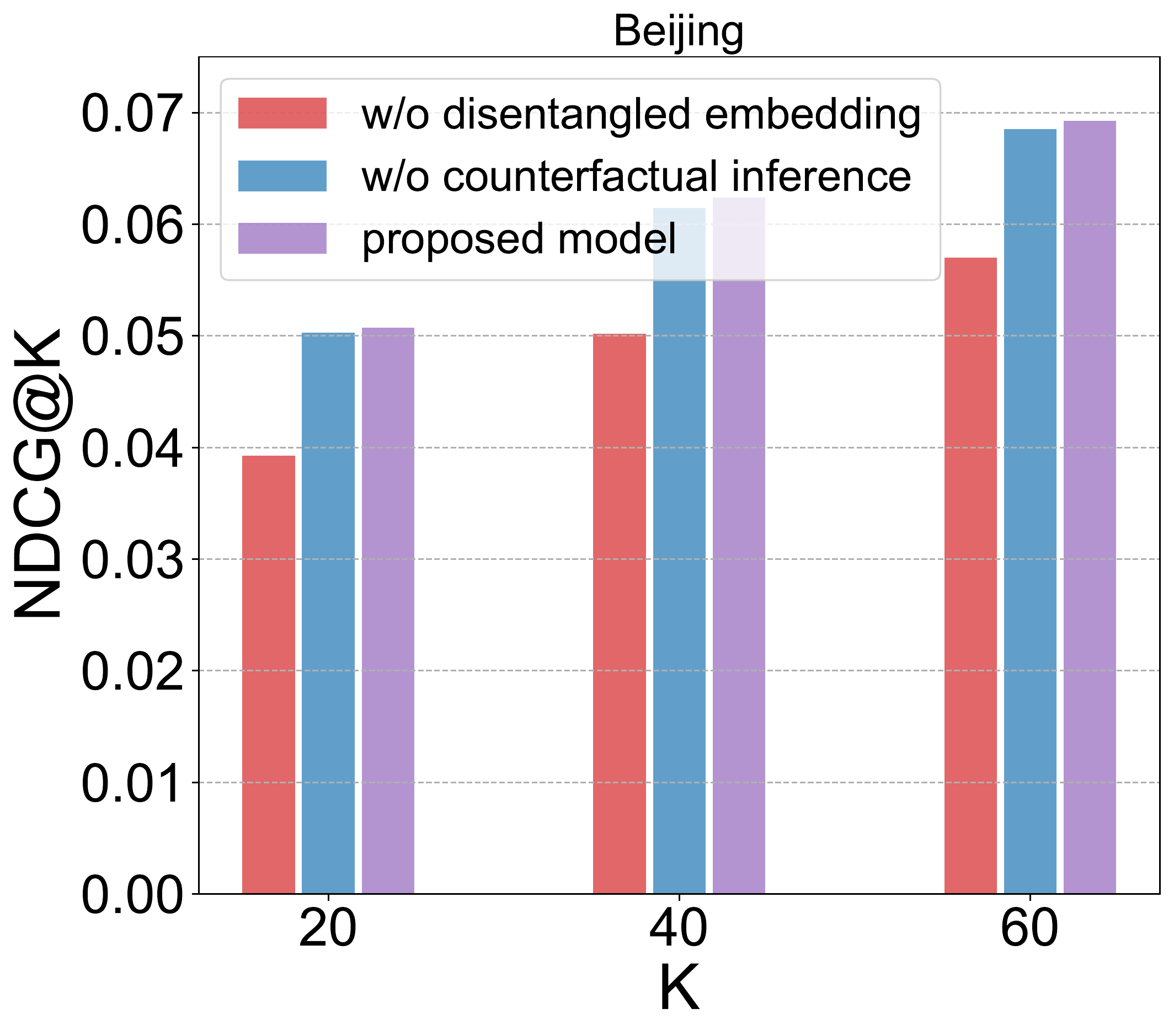}}
    \vspace{-0.4cm}
    \caption{Ablation study on Beijing dataset.}
    \label{fig:ablationbj}
    \vspace{-0.4cm}
\end{figure}
\begin{figure}[t]
    \centering
    \vspace{-0.2cm}
    \subfigure[Recall.]{
    \label{abrecallsh}
    \includegraphics[width=0.47\linewidth]{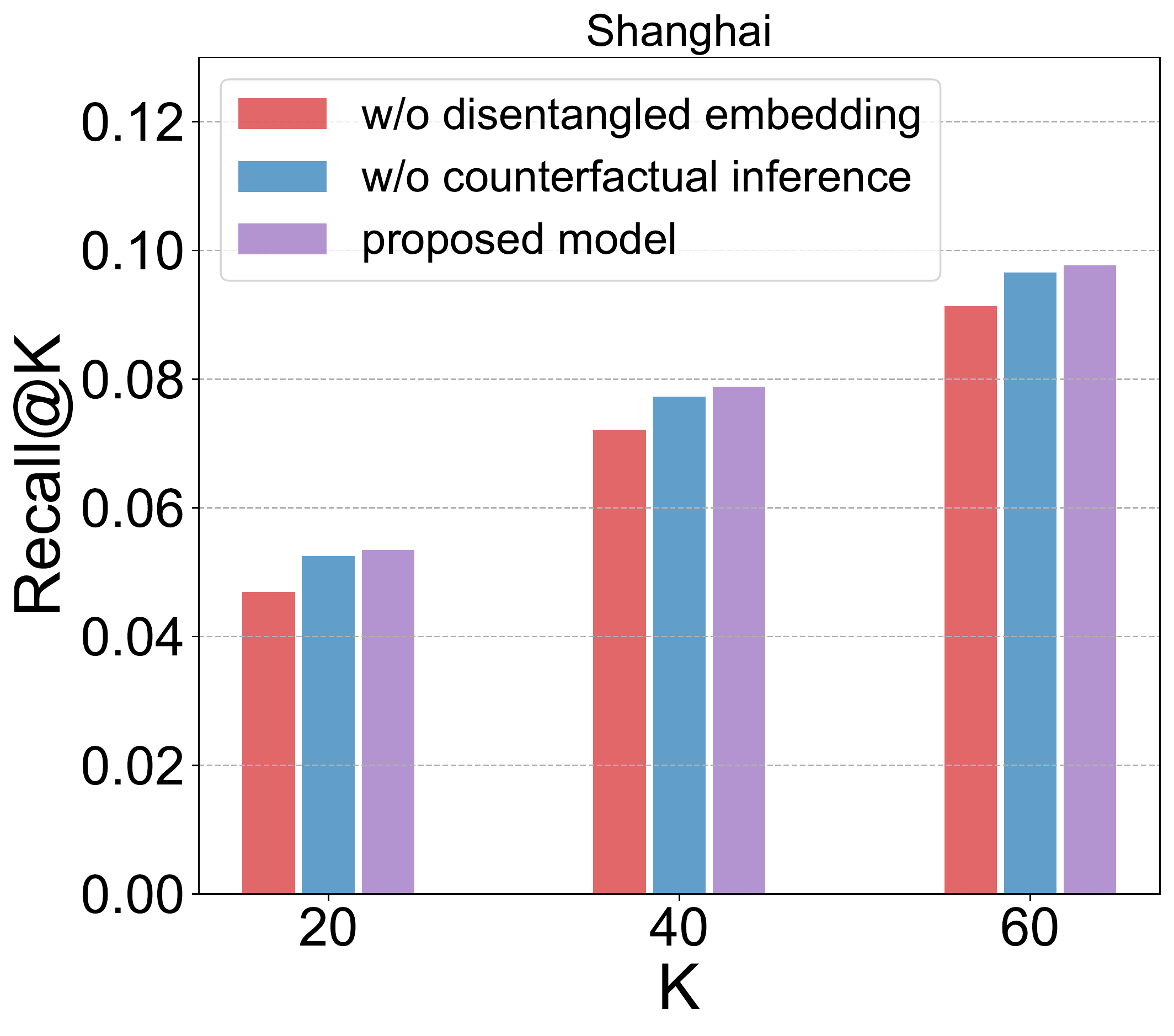}}\subfigure[NDCG.]{
    \label{abndcgsh}
    \includegraphics[width=0.47\linewidth]{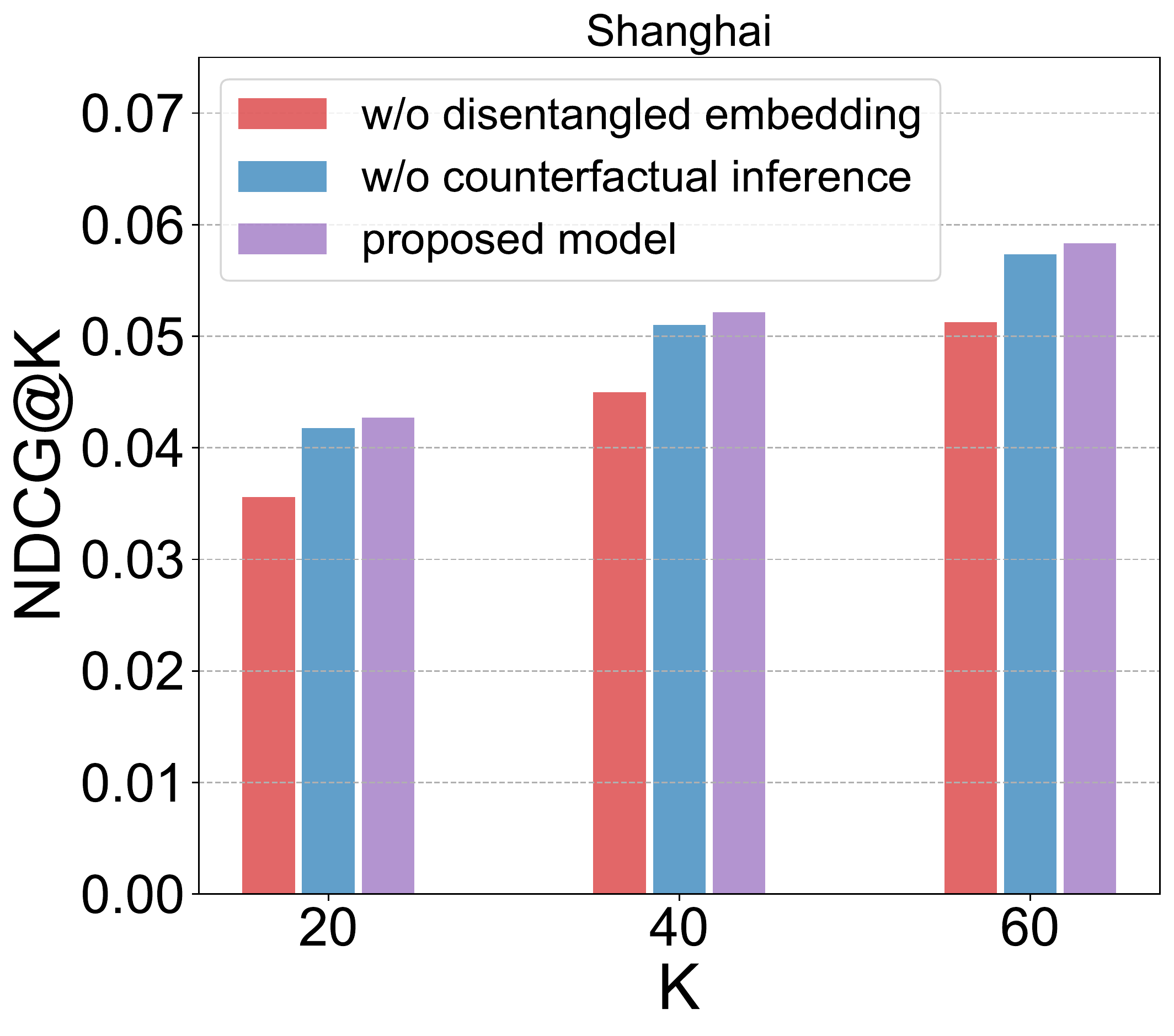}}
    \caption{Ablation study on Shanghai dataset.}
    \label{fig:ablationsh}
\end{figure}
We perform ablation experiments to measure each designed components.
In our UKGC method, we first deploy separated information propagation layers on the disentangled embeddings
and then adopt counterfactual inference for prediction.
To study the effectiveness of these two parts of designs, we remove one of them and test the performance.
To remove the design of disentangled embeddings,
we learn two sets of embeddings with two complete urban knowledge graphs without dividing.
Precisely speaking, we give each set of embeddings blended knowledge to learn the representations and rating with counterfactual learning, to see whether it can capture geographical and functional attributes under this situation. 
To remove the design of counterfactual inference, different from (\ref{fml::tie}), we just use $\mathrm{TE} = \mathrm{Y}_{u_i,p_j,g_j} - \mathrm{Y}_{u_i, p^{*}_{j}, g^{*}_{j}}$ as predicted $\hat{y}_{u_i,p_j}$. 

The experimental results of Beijing and Shanghai datasets are reported in Figure~\ref{fig:ablationbj} and~\ref{fig:ablationsh}, respectively. 
\begin{itemize}
    [leftmargin=*,partopsep=0pt,topsep=0pt]
    \setlength{\itemsep}{0pt}
    \setlength{\parsep}{0pt}
    \setlength{\parskip}{0pt}
    \item Without the design of disentangled embedding, the performance are much worse than original proposed model on both two datasets. Specifically, the relative decrease is $11.3\%-17.3\%$ in Recall and $17.7\%-22.6\%$ in NDCG in Beijing dataset, $6.4\%-12.1\%$ in Recall and $13.7\%-16.6\%$ in NDCG in Shanghai dataset.
    It indicates the importance of knowledge disentanglement, which is essential for the better usage of the urban knowledge graph.
    \item Without counterfactual inference rating, the performance on all metrics also has a certain degree of decrease in both dataset. Specifically, the relative decrease is $1.0\%-2.0\%$ in Recall and $1.0\%-1.5\%$ in NDCG in Beiijing dataset,  $1.1\%-1.9\%$ in Recall and $1.7\%-2.3\%$ in NDCG in Shanghai dataset.
    It demonstrates the importance of counterfactual inference to alleviate geographical bias. 
\end{itemize}

\vspace{-0.2cm}
\subsection{Hyper-parameter Study (RQ3)}
In this section, we explore the effect of different hyper-parameters on model performance.
We conduct extensive studies of 1) the number of propagation layers $l$ in (\ref{fml::players})~and~(\ref{fml::users}), 2) the relative weight $\alpha$ in (\ref{fml::totalloss}), 3) the number of user intents $|\mathcal{I}|$ in~(\ref{fml::int}), and 4) fusion strategy in ~(\ref{fml::mt}). 
\textbf{Effect of $\alpha$.}
\begin{figure}[t]
    \centering
    \subfigure[Recall.]{
    \label{alpharecall}
    \includegraphics[width=0.47\linewidth]{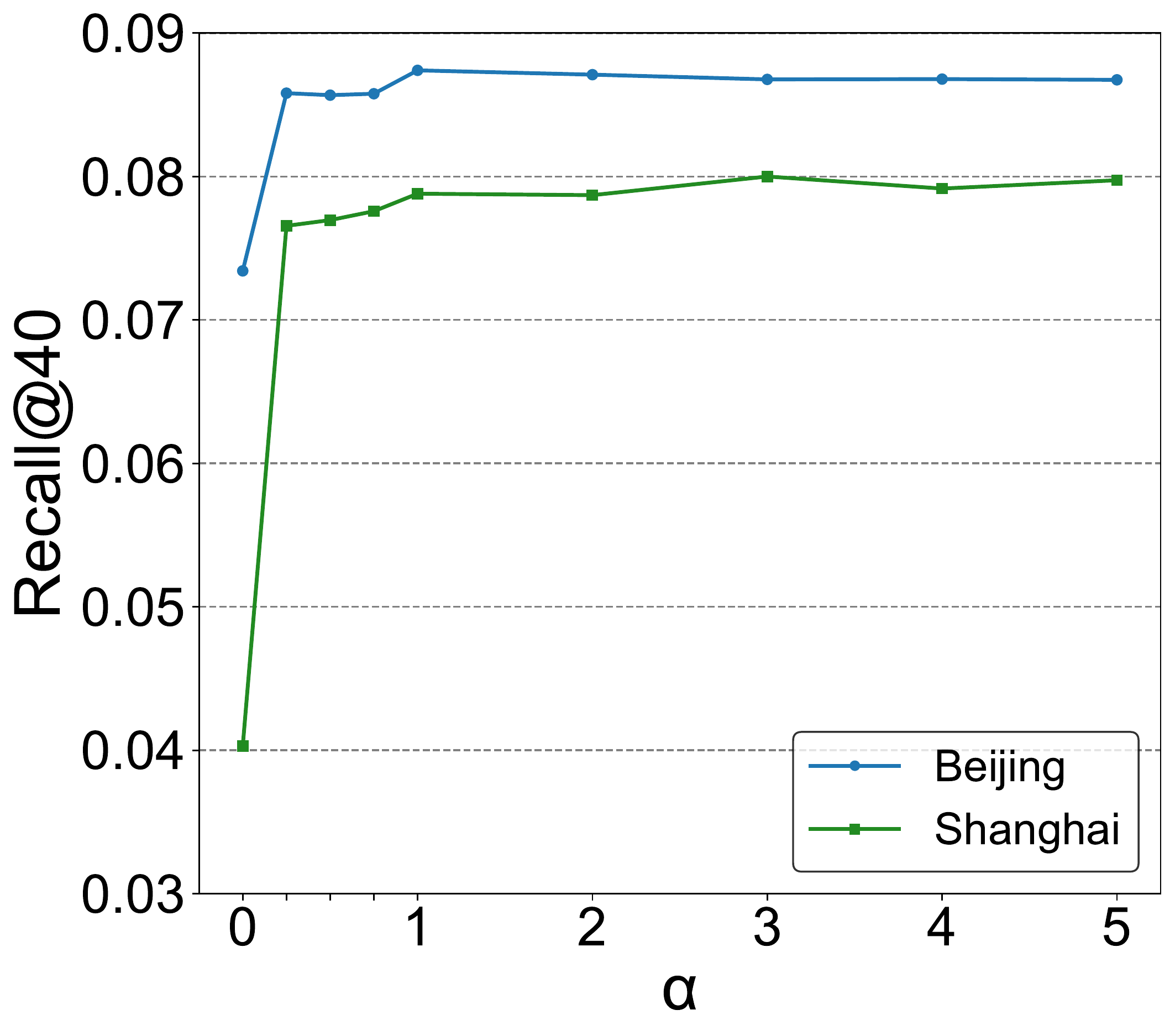}}\subfigure[NDCG.]{
    \label{alphandcg}
    \includegraphics[width=0.47\linewidth]{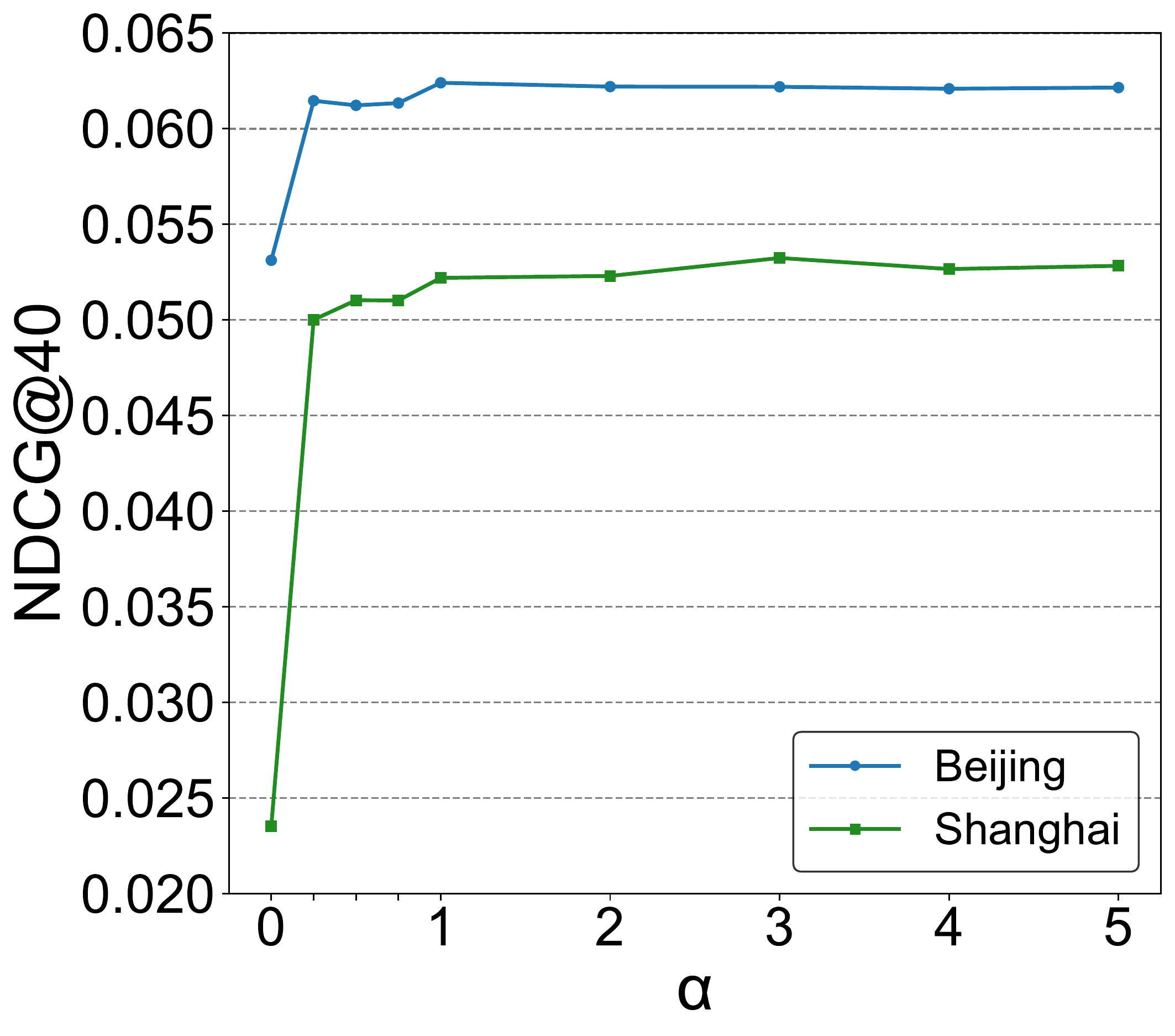}}
        \vspace{-0.4cm}
    \caption{Performance of different $\alpha$ in Beijing dataset.}
    \label{fig:alpha}
        \vspace{-0.4cm}
\end{figure}
Hyper-parameter $\alpha$ represents the relative weight of learning in the factual world and the counterfactual world. 
Since different datasets have different geographical bias, $\alpha$ needs to be tuned. The results of different $\alpha$ values are shown in Figure~\ref{fig:alpha}.
We can observe that setting $\alpha=0$ makes performance worse, reflecting that we cannot abandon learning in the counterfactual world. 
As $\alpha$ rises, the performance of the model also rises gradually, illustrating the importance of learning in the counterfactual world. 
The experimental results show that $\alpha=1$ is appropriate on Beijing dataset, while $\alpha=3$ seems better to Shanghai dataset. 
Further making $\alpha$ continue to increase does not result in a significant improvement for model performance.\\
\textbf{Effect of propagation layers.}
\begin{figure}[t]
    \centering
    \subfigure[Recall.]{
    \label{hopsrecall}
    \includegraphics[width=0.47\linewidth]{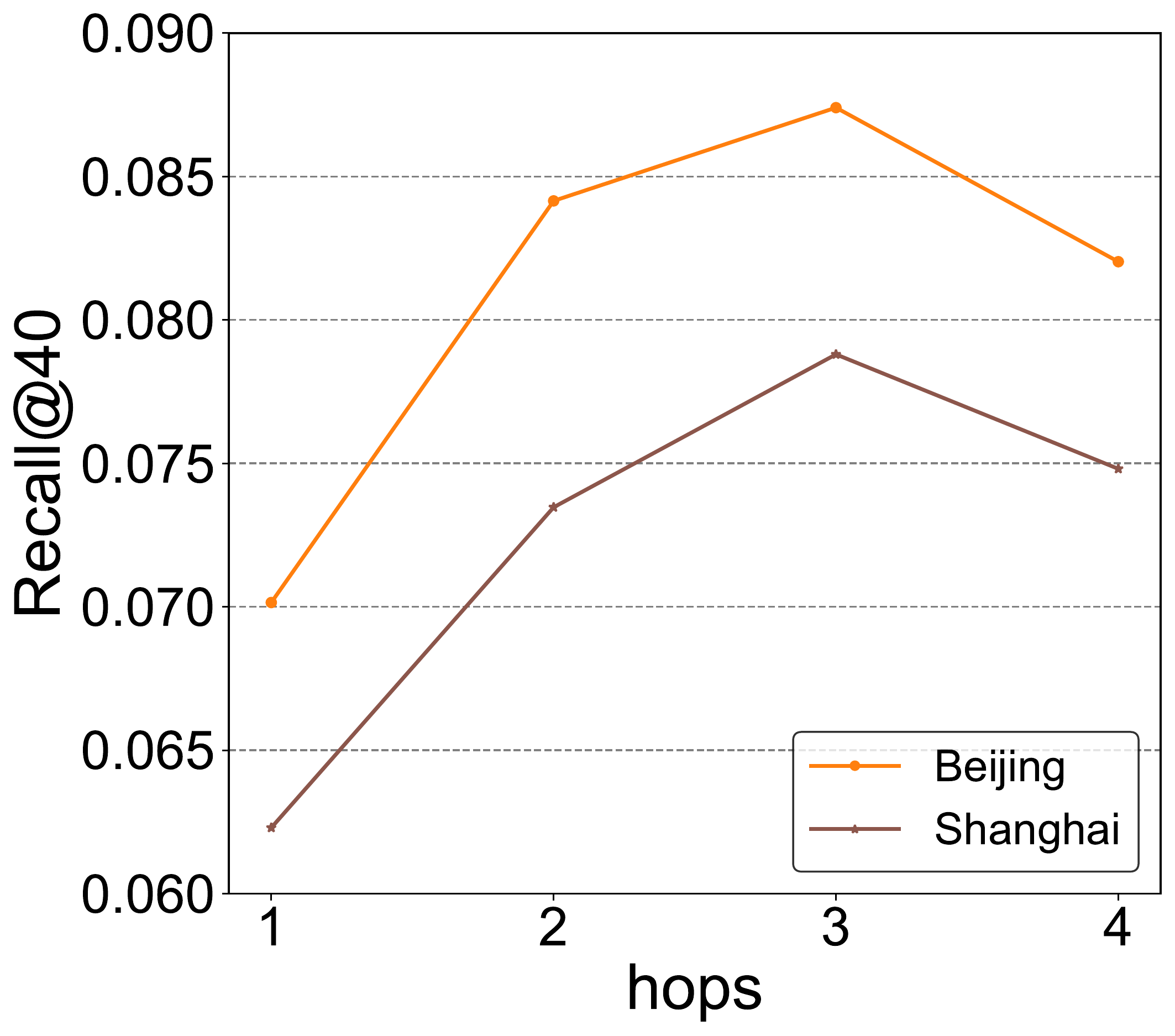}}\subfigure[NDCG.]{
    \label{hopsndcg}
    \includegraphics[width=0.47\linewidth]{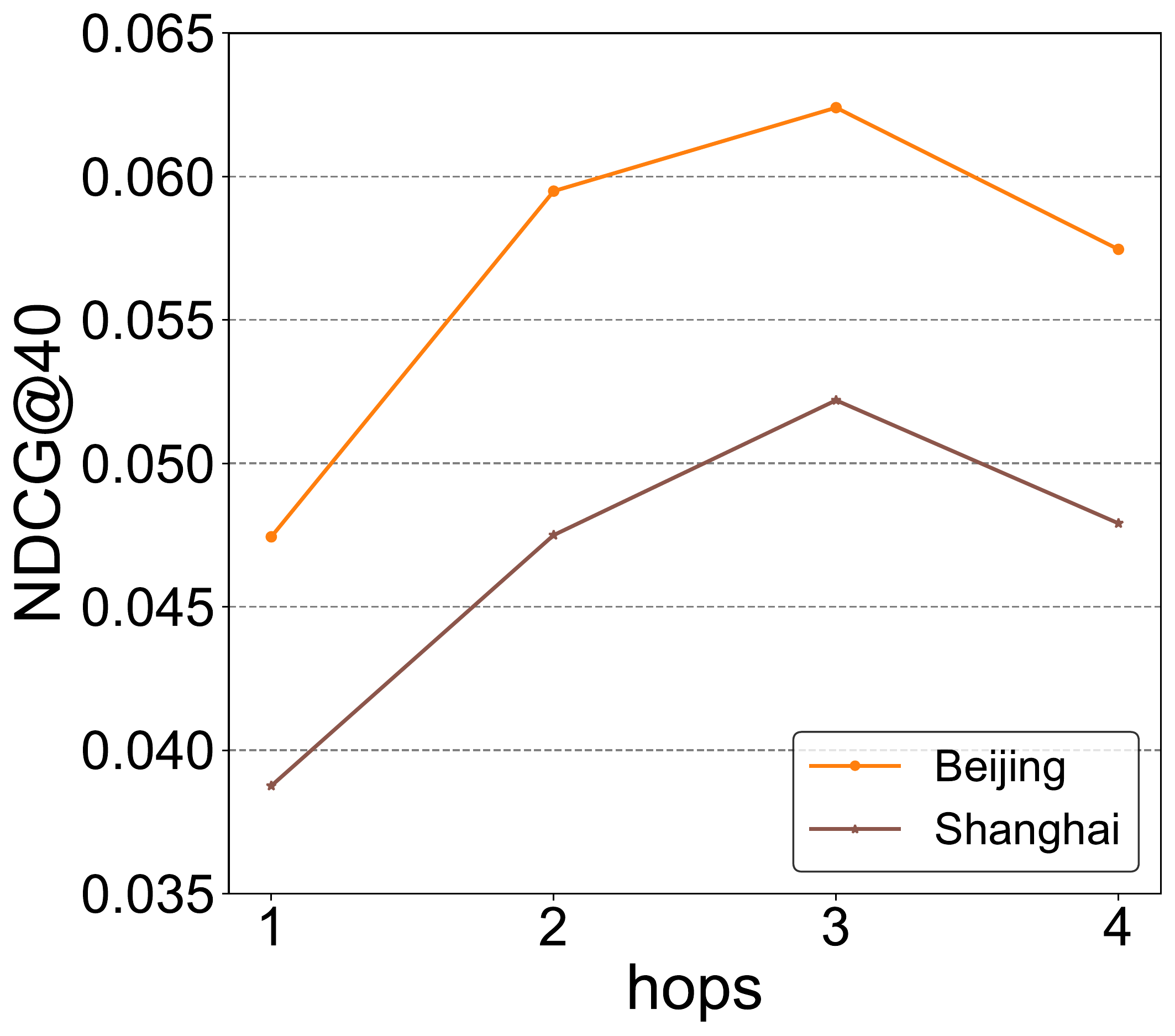}}
        \vspace{-0.4cm}
    \caption{Performance of different propagation layers.}
        \vspace{-0.4cm}
    \label{fig:hops}
\end{figure}
We also test the effect of different numbers of propagation layers, also called \textit{hops}.
Following existing works~\cite{wang2019kgat,wang2021kgin}, we search hops in $\{1,2,3,4\}$ to find the most proper number of layers. 
The results are reported in Figure~\ref{fig:hops}. 
We find that our model can reach the best performance with three propagation layers. 
Fewer propagation layers can only obtain information from low-order connectivity, i.e. their nearest neighbors, which limits learning better representations.
On the other hand, stacking too many propagation layers could make all nodes aggregate information from almost all other nodes in the graph, which may make the representations of some nodes indistinguishable and over-fitting. 
Hence, we recommend three propagation layers in our model.\\
\textbf{Effect of the number of user intents.}
\begin{figure}[t!]
    \centering
    \subfigure[Recall.]{
    \label{intentsrecall}
    \includegraphics[width=0.47\linewidth]{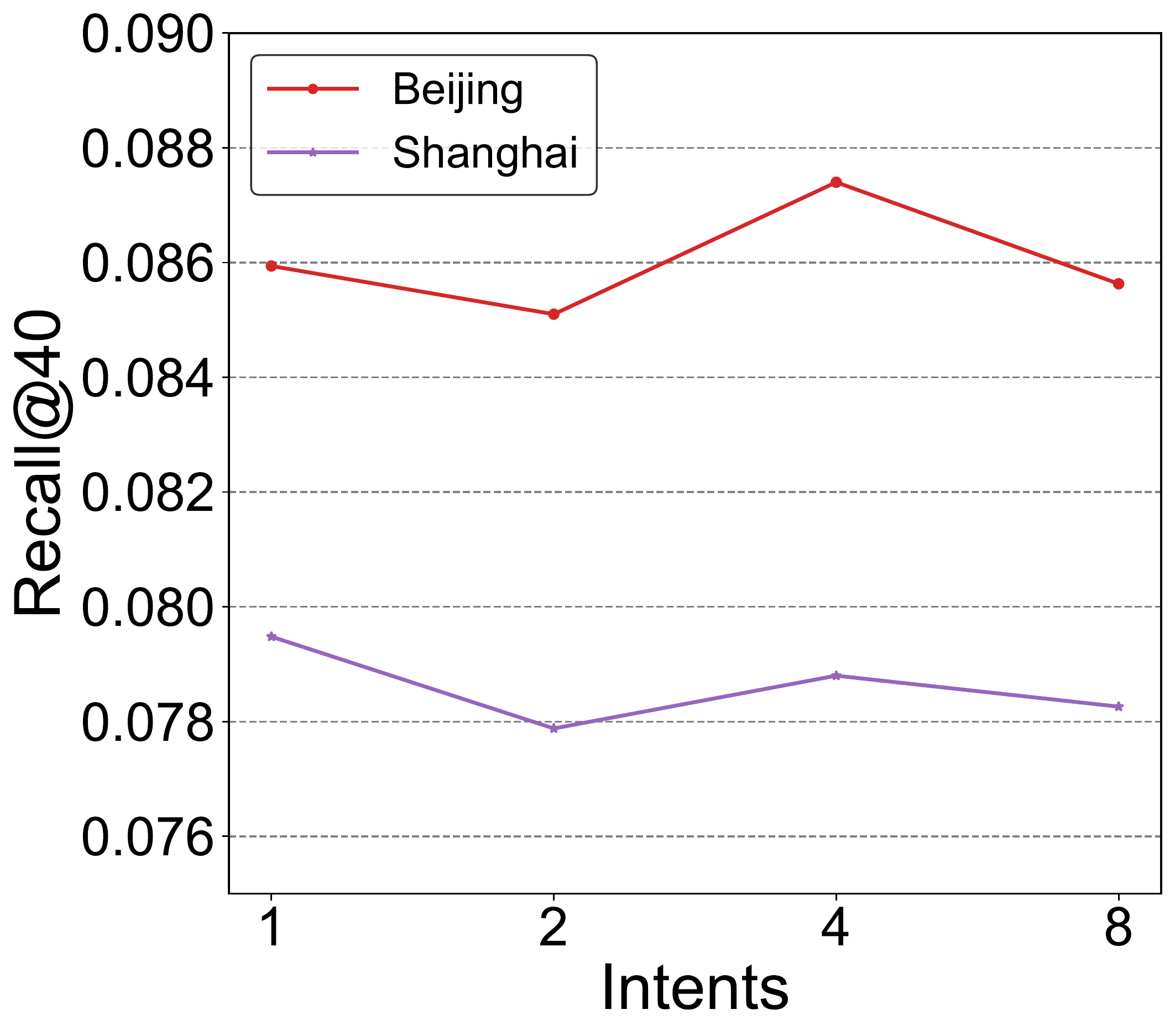}}\subfigure[NDCG.]{
    \label{intentsndcg}
    \includegraphics[width=0.47\linewidth]{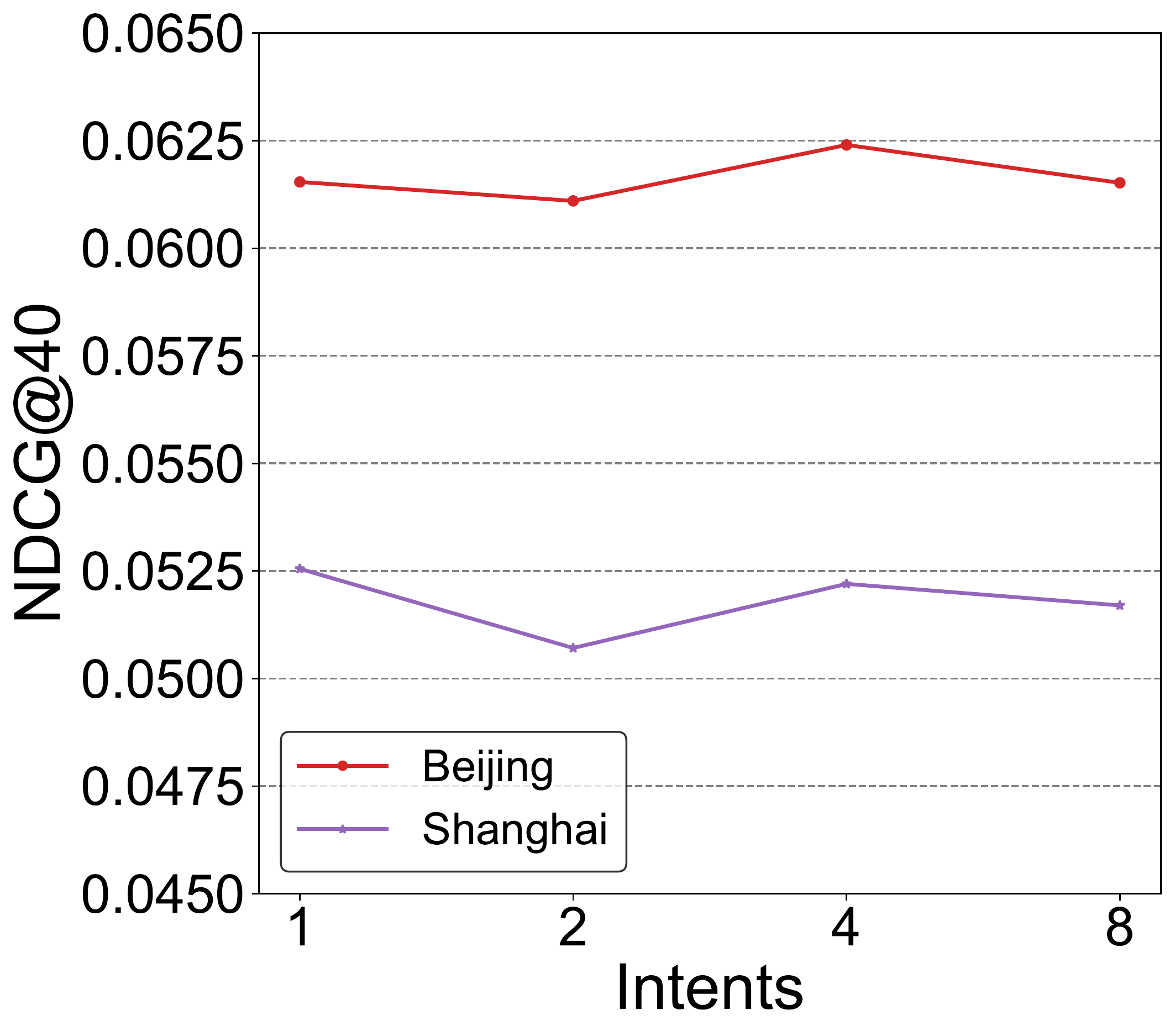}}
    \vspace{-0.4cm}
    \caption{Performance of different user intents.}
    \label{fig::int}
    \vspace{-0.4cm}
\end{figure}
We also analyze the influence of the number of user intents in each partitioned graph. We search the number of intents $|\mathcal{I}|$ in $\{1,2,4,8\}$ and present the results in \text{Figure \ref{fig::int}}. We find that the number of user intents needs to be adjust for different datasets due to the heterogeneity of the data. For example, we can obtain the best performance on Beijing dataset by setting $|\mathcal{I}|=3$, however $|\mathcal{I}|=1$ seems to be a better choice for Shanghai dataset. Too much independent intents may be difficult to carry useful information, which could explain the decrease of performance when the number of intents becomes large.
\vspace{-0.4cm}
\subsection{Case Study (RQ4)}
\begin{figure}[t]
    \centering
    \subfigure[Beijing.]{
    \label{bjcase}
    \includegraphics[width=0.48\linewidth]{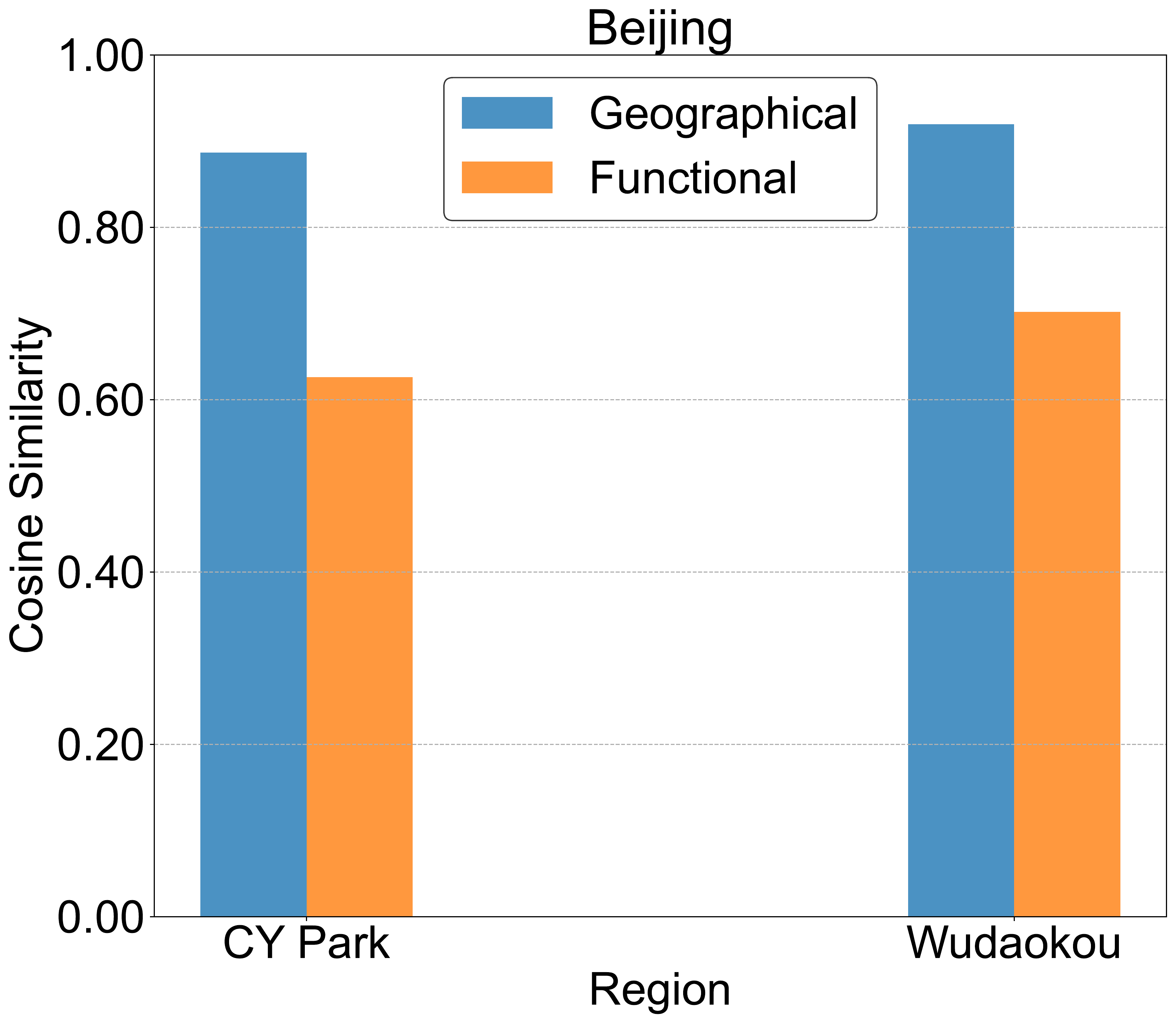}}\subfigure[Shanghai.]{
    \label{shcase}
    \includegraphics[width=0.48\linewidth]{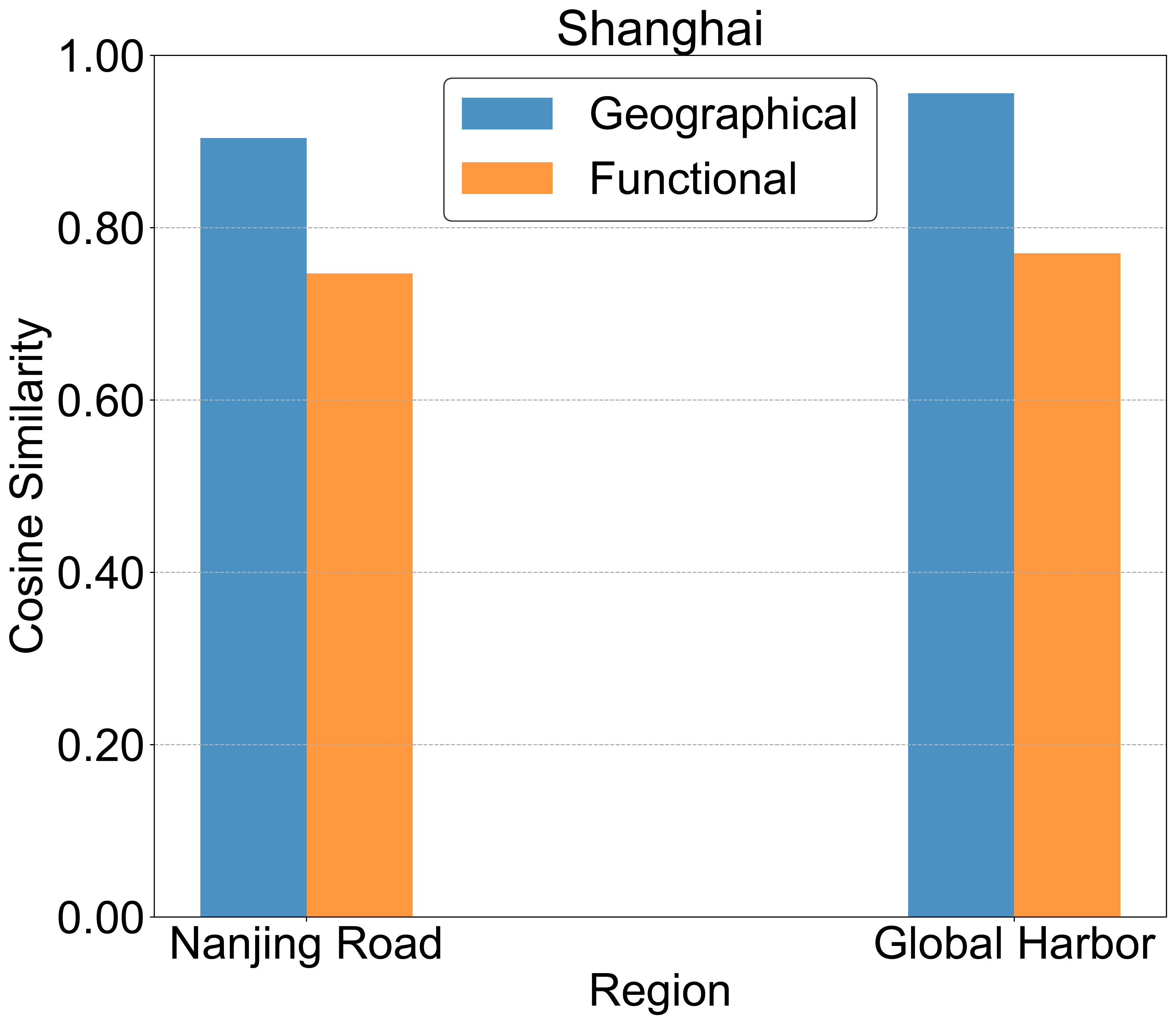}}
        \vspace{-0.4cm}
    \caption{The cosine similarity of two parts of embeddings.}
    \label{fig::sml}
        \vspace{-0.4cm}
\end{figure}
\vspace{-0.1cm}
To further investigate POIs' embedding learning of geographical and functional aspects and verify the effect of disentangling, we choose two regions in each city with 
enough number of POIs and interactions, 
then observe the similarity of geographical and functional embeddings of POIs located in the same region. 
CY Park and Wudaokou are chosen in Beijing, which are two famous regions with diverse POIs. We select Nanjing Road and Global Harbor in Shanghai for the same reason. 
We then calculate the cosine similarity of every two POIs' disentangled embeddings and use their average values as the cosine similarity of the geographical or functional embeddings of POIs in the region. The results are presented in the form of a bar chart in Figure~\ref{fig::sml}.

From the figure, we can know that the geographical embeddings of POIs are significantly more similar compared with their functional embeddings within the selected regions, which is in line with our expectations. 
Specifically, POIs in the same region have similar geographical attributes while their functional attributes should be different. Experimental results validate our model can capture geographical and functional aspects well, thus demonstrating that we do learn the representation of POIs more accurately.

	\section{Related Work}\label{sec::relatedwork}
\textbf{Point-of-Interest Recommendation. }
In recent years, various methods have been proposed to improve the performance of POI recommendation tasks~\cite{Lim2020STP,chang2020GPR,chang2018content,lian2014geomf}. In addition to the interaction data between users and POIs, side information has been widely used. For example, RankGeoFm~\cite{li2015rank} generates two embeddings for each user, representing users' preference and nearby POIs respectively. MTEPR~\cite{chen2021multi} leverages social relations and profiles of users, categories and locations of POIs for recommendation tasks, integrating them into multiple bipartite graphs. IRenMF~\cite{liu2014exploiting} exploits the similarity between POIs according to the geographical distance. In our work, we construct the UrbanKG containing a wide variety of side information, which is more comprehensive and strutured than the previous models. We think it can facilitate the learning of representations of users and POIs.\\
\textbf{Knowledge Graph-based Recommendation. }
Before we introduce the urban knowledge graph into the field of POI recommendation, the knowledge graph has been used in other recommendation tasks~\cite{ai2018learning,zhang2016cke,cao2019unifying}. 
For example, CKE~\cite{zhang2016cke} exploits TransE to obtain embeddings of entities in the KG, feeding them into matrix factorization. CFKG~\cite{zhang2018cfkg} regards interaction between users and POIs as a kind of relation in the KG and 
learns to predict whether this relation exists.
Other methods~\cite{hamilton2017inductive,wang2019kgcn,wang2020ckan,Wang2019ngcf} utilize information aggregation mechanism of graph neural networks to propagate information on the KG.
For example, KGAT~\cite{wang2019kgat} propagates on the composite graph intergrating KG and interactions, adopting attention mechanism. In addition to information propagation on the KG, KGIN~\cite{wang2021kgin} models the intent layer between users and POIs. To our best knowledge, our work is the first attempt to introduce the urban knowledge graph to the POI recommendation task.\\
\textbf{Causal Recommendation. }
Causal recommendation is often used to debias in recommendation tasks~\cite{Joa2017unbiased,christakopoulou2020deconfounding,zhu2020unbiased}. For example, IPW~\cite{liang2016causal} estimate the propensity of popular, and re-weight samples using the inverse propensity scores. MACR~\cite{wei2021model} conducts a multi-task learning with counterfactual inference for eliminating the popularity bias. Wang et al~(\citeyear{wang2021clicks}) use counterfactual inference to eliminate exposure bias. Based on IPS, McInerney et al~(\citeyear{McInerney_2020}) proposes Reward Interaction IPS to eliminate the bias by logged data when evaluating offline. 
In our work, we introduce counterfactual inference to remove the geographical bias in interactions and achieve better performance.
	\section{Conclusion and Future Work}\label{sec::conclusion}
In this work, we address the POI recommendation from a brand new perspective, taking a pioneering step to introduce the urban knowledge graph and combine the counterfactual inference to this problem.
We first build the urban knowledge graph that contains structural information of POIs, which can be disentangled into the functional aspect and the geographical aspect.
Since the geographical factor plays as a confounder, we propose a graph neural network-based model, UKGC, with counterfactual learning.
Extensive experiments verify the effectiveness of the constructed UrbanKG and the proposed UKGC method.
As for the future work, we plan to conduct online A/B tests in real-world recommender systems, to further evaluate the effectiveness of the UrbanKG and the UKGC model.
	\bibliographystyle{ACM-Reference-Format}
	\bibliography{bibliography}

\end{document}